\documentclass[aps,prb,twocolumn,showpacs]{revtex4}
\usepackage{graphicx}
\usepackage{amsmath}
\usepackage{amsbsy}
\usepackage{mathbbol}
\usepackage{color}
\graphicspath{{figures/}}

\newcommand{\HM}{H\,}
\newcommand{\eh}{\mathrm{he}}
\newcommand{\he}{\mathrm{eh}}
\newcommand{\lat}{\mathrm{eff}}
\renewcommand{\vr}{\mathbf{r}}
\renewcommand{\vec}[1]{\mathbf{#1}}
\newcommand{\pot}{{\mu}}
\newcommand{\vp}{\vec{p}}
\newcommand{\vq}{\vec{q}}
\newcommand{\vk}{\vec{k}}
\newcommand{\vkp}{\vec{k}_{\parallel}}
\newcommand{\vkprp}{\vec{k}_{\perp}}
\newcommand{\up}{\uparrow}
\newcommand{\down}{\downarrow}
\newcommand{\vOmega}{\boldsymbol{\Omega}}

\begin{document}

\title{Andreev reflection from non-centrosymmetric superconductors and
Majorana bound state generation in half-metallic ferromagnets}

\author{Mathias Duckheim}

\author{Piet W. Brouwer}

\affiliation{Dahlem Center for Complex Quantum Systems and
Institut f\"ur Physik, Freie Universit\"at Berlin, D-14195 Berlin, Germany}


\date{\today}

\begin{abstract} 
We study Andreev reflection at an interface between a half metal and
a superconductor with spin-orbit interaction. While the absence of
minority carriers in the half metal makes singlet Andreev reflection
impossible, the spin-orbit interaction gives rise to triplet Andreev
reflection, i.e., the reflection of a majority electron into a
majority hole or vice versa.  As an application of our calculation, we
consider a thin half metal film or wire laterally attached to a
superconducting contact. If the half metal is disorder free, an
excitation gap is opened that is proportional to the spin-orbit
interaction strength in the superconductor. For electrons with energy
below this gap a lateral half-metal--superconductor contact becomes a
perfect triplet Andreev reflector. We show that the system supports
localized Majorana end states in this limit.
\end{abstract}
\maketitle

\section{Introduction}

Heterosystems with adjacent superconducting and ferromagnetic phases
may show unconventional spin-triplet superconducting proximity effects
even if the superconductor is of the conventional $s$-wave
spin-singlet type.\cite{Bergeret2005} Triplet correlations, even if
they are weak, are important in ferromagnets, where the standard
spin-singlet proximity effect is short-ranged 
as a result of the exchange splitting.\cite{Buzdin2005} In half-metals
singlet pairings are ruled out, since in a half metal one spin species
has zero density of states at the Fermi level, so that the triplet
version is the only possible form of the superconducting proximity
effect. Microscopically, the superconducting proximity effect is
mediated by Andreev reflection, the phase-coherent reflection of an
electron into a hole or vice versa at the superconductor
interface.\cite{Andreev1964} Triplet superconducting correlations then
require a form of Andreev reflection that includes a spin
flip.\cite{Bergeret2001,Kadigrobov2001}

The triplet proximity effect has been considered first in ferromagnets
(with a finite minority spin population), where the observation of
long-range superconducting correlation
effects\cite{Petrashov1999,Sosnin2006,Krivoruchko2007,Yates2007,Khaire2010}
has been shown to be consistent with the existence of induced triplet
correlations in the ferromagnet.\cite{Bergeret2001,Bergeret2005} A
number of mechanisms that give rise to the spin-flip Andreev
reflection required for the triplet correlations, such as magnetic
domain walls,\cite{Bergeret2001} spin-orbit interaction,\cite{Wu2009}
and unconventional pairing correlations,\cite{Linder2007} have been
studied in hybrid ferromagnet-superconductor systems.

\begin{figure}
\flushleft a \\
\begin{minipage}[l]{1.0\linewidth}
  \includegraphics[width = 0.75 \linewidth]{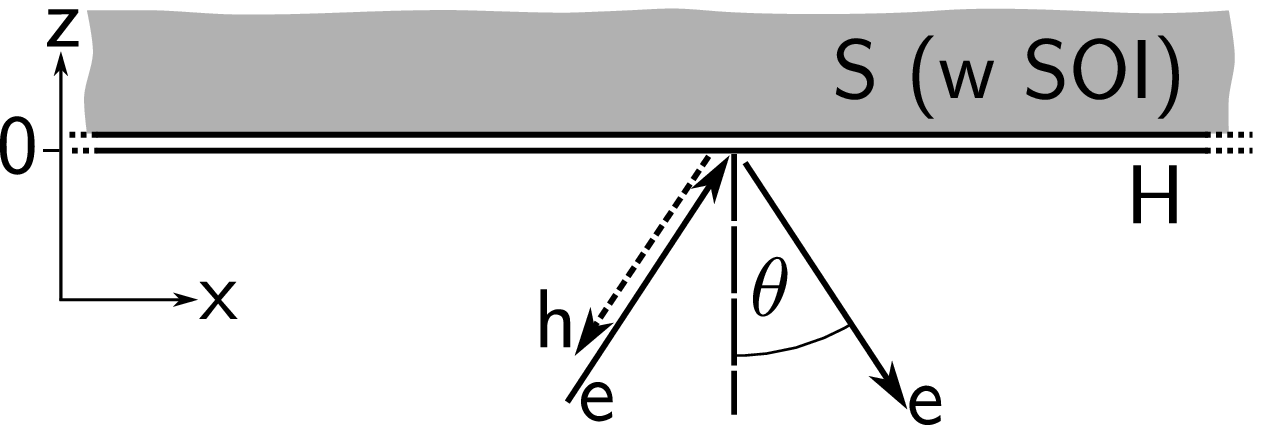}\\
\end{minipage}
  
\bigskip
b \\
\begin{minipage}[b]{1.0\linewidth}
  \includegraphics[width = 0.98 \linewidth]{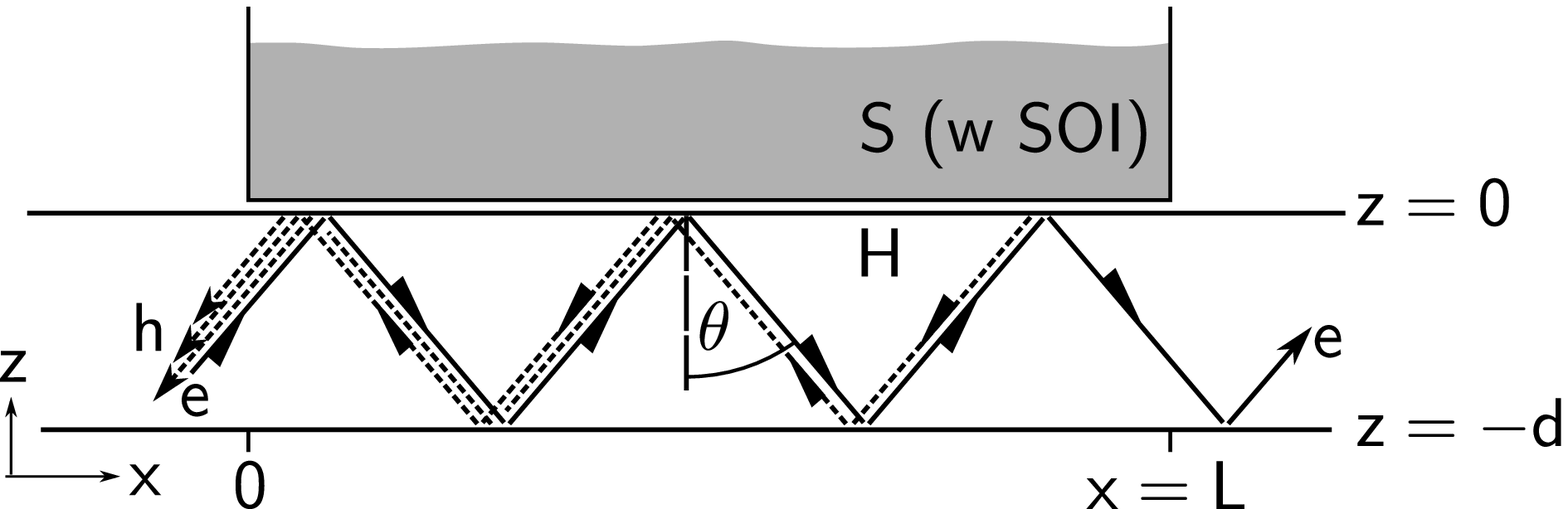}
\end{minipage}
\caption{Top panel (a): Interface between a half metal (\HM) and a
  superconductor (S). For a superconductor with spin-orbit scattering,
  normal reflection as well as triplet Andreev reflection --- the
  reflection of a majority electron (e) into a majority hole (h) ---
  take place at the HS interface. Bottom panel (b): A lateral contact
  between a half metallic film or wire of thickness $d$ and a
  superconductor. For a lateral contact, an effective Andreev
  reflection amplitude $r^{\mathrm{\lat}}_{\eh}$ for electrons moving
  towards the contact region can be defined, which describes the
  combined effect of multiple reflections at the \HM-S interface. }
  \label{fig:HM-S-interface}
\end{figure}

Long-range proximity effects have also been observed in half-metallic
$\mathrm{CrO_2}$--NbTiN\cite{Keizer2006} and
$\mathrm{CrO_2}$--MoGe\cite{Anwar2010} heterostructures. Because of
the absence of minority carriers in the half metal it is concluded
that the observed superconducting correlations must be of triplet
type.\cite{Keizer2006,Eschrig2008} However, in half metals the
conditions for spin-flip Andreev reflection are more restrictive than
in a ferromagnet. In particular, electron-hole symmetry and current
conservation pose stronger restrictions on candidate mechanisms for
spin-flip Andreev reflections than in
ferromagnets:\cite{beri2009,Kupferschmidt2010} If the interface is
symmetric with respect to reflection in the surface normal,
electron-hole symmetry and unitarity require the Andreev reflection
amplitude $r_{\eh}(\varepsilon)$ to vanish at the Fermi
energy $\varepsilon =0$.\cite{beri2009} Thus, mechanisms that give rise to triplet
Andreev reflection must either break electron-hole symmetry, {\em
  i.e.} take place away from the Fermi energy, or orbital
symmetries. Mechanisms of the latter type are magnetization gradients
in the half metal,\cite{Kupferschmidt2009,Kupferschmidt2010} or
impurity scattering.\cite{Wilken2010}

In this article, we study spin-orbit interaction (SOI) in S as a
possible mechanism giving rise to spin-flip Andreev reflection in a
half metal (\HM). The presence of SOI is contingent on the breaking of
inversion symmetry. Examples for systems with SOI in the normal state
are surface states in Au,\cite{LaShell1996} semiconductor
heterostructures and 2D electron gases in quantum wells with partially
tunable SOIs.\cite{rashba:1960a,Nitta1997,Kato2004} Crystalline
superconductors such as the non-centrosymmetric cuprates and
heavy-fermion compounds such as $\mathrm{CePt_3Si}$\cite{Bauer2004}
and others,\cite{Sigrist2007} show SOI due to the absence of an
inversion center in their crystal structure. These materials have
received intense interest because they display unconventional
superconducting (helical\cite{Agterberg2003,Samokhin2004,Kaur2005})
phases with mixed singlet and triplet pairing
correlations,\cite{Sigrist1991,Frigeri2004,Nelson2004,Sigrist2007}
magnetoelectric effects,\cite{Edelstein1995,Fujimoto2005} and an
anisotropic spin susceptibility.\cite{Gorkov2001,Yip2002} But also
centrosymmetric superconducting materials show SOI in surface or
interfacial layers with inversion asymmetry. For instance, the
breaking of inversion symmetry at a plane interface due to a change in
the chemical potential gives rise to the Rashba
SOI.\cite{rashba:1960a,Gorkov2001}

In our calculation we assume a general model for the SOI that is
linear in momentum and includes the Rashba SOI as a special case. In
the first part of this article, we derive expressions for the
electron-to-hole and hole-to-electron Andreev reflection amplitudes
$r_{\eh}$ and $r_{\he}$ in a model Hamiltonian with parabolic
dispersion in the half metal and in (the normal state of) the
superconductor and to first order in the SOI. For our model
Hamiltonian, which includes effects of a Fermi-velocity mismatch and a
tunnel barrier at the half-metal--superconductor interface, we find
that the Andreev reflection is such that the induced superconducting
correlations in the vicinity of the superconducting interface are of
even-frequency and complex $p$-wave type. This is a significant
difference with other possible sources of triplet superconducting
correlations in half metals, such as a nonuniform magnetization
direction in the half metal, which also allow for odd-frequency
$s$-wave proximity effects.

Because Andreev reflection relies on the presence of SOI in the
superconductor, the Andreev reflection probability $|r_{\eh}|^2$ may
be small depending on the strength of the SOI such that the induced
superconducting correlations become weak. However, a fully developed
proximity effect\cite{Braude2007} can be achieved in a geometry in
which multiple Andreev reflections occur. Examples of such geometries
are a half-metallic film or wire in contact to a superconductor (shown
schematically in Fig.~\ref{fig:HM-S-interface} b). The latter example is
closely related to recent proposals for the realization of Majorana
fermions in a solid-state setting,\cite{Kitaev2000,kn:sau2010,
  Lee2009-pa,Lutchyn2010, Oreg2010, Potter2010,Alicea2010} which in
turn play an important role in proposals for topological quantum
computation.\cite{Kitaev2000,Nayak2008} The second part of this
article contains an investigation of multiple Andreev reflections in
the film- or wire geometry. We will show that the effect of multiple
Andreev reflections can be combined into effective Andreev reflection
amplitudes $r_{\eh}^{\lat}$ and $r_{\he}^{\lat}$. These effective
amplitudes may have unit magnitude if sufficient reflection events
contribute coherently them. We show that in this case localized
Majorana states can be formed at the ends of a half metal wire in
contact to a superconductor with spin-orbit coupling. The
investigation of the thin-film geometry is motivated by the recent
experiments of Refs.\ \onlinecite{Keizer2006,Anwar2010}.

In the present calculation we do not consider scattering from
impurities in \HM or S. This does not seriously affect the first part
of our calculation, which addresses the Andreev reflection amplitude
for a single reflection off the HS interface, because the Andreev
reflection amplitude is a local property of the interface. It is,
however, a limitation for the second part of our calculation, since
the proximity effect induced by the multiple Andreev reflections is of
$p$-wave type, which is suppressed by impurity scattering. Thus, the
results derived in the second part of this article are valid only if
the elastic mean free path in the half metallic film or wire is
sufficiently large.

The article is structured as follows. In Sec.~\ref{sec:model}, we
introduce the model Hamiltonian of the HS heterostructure and describe
the calculation of the Andreev reflection amplitudes for a single
reflection off a HS interface. As an application of our calculation,
we calculate the subgap conductance of a half-metal--superconductor
interface and the Josephson current in a
superconductor--half-metal--superconductor junction in Sec.\
\ref{sec:subg-cond-josephs}. In Sec.~\ref{sec:lateral-geometry} we
consider the geometry in which a thin half-metallic film is brought
into electrical contact with a superconductor and derive the effective
Andreev reflection amplitude $r_{\eh}^{\lat}$ for this
situation. Applications to the subgap conductance and Josephson effect
in lateral HS and SHS junctions are then given in Sec.\
\ref{sec:subg-cond-josephs-1}. Finally, in
Sec.~\ref{sec:majorana-states}, we consider a half-metallic wire
placed in contact to a superconductor and relate our findings to
predictions of the occurrence of Majorana fermions in such a system.

\section{Interface between half metal and superconductor}\label{sec:interf-betw-half}

\subsection{Hamiltonian}

\label{sec:model}

We consider the interface between a half metal (\HM) and a
superconductor (S) as shown in Fig.~\ref{fig:HM-S-interface}
a. Coordinates are chosen, such that the superconductor and the half
metal occupy the half spaces $z > 0$ and $z < 0$,
respectively. Electron and hole excitations at excitation energy
$\varepsilon$ near the interface are described by the Bogoliubov-de
Gennes equation
\begin{align}
  \label{eq:BdG-eq}
  \mathcal{H} \Psi = \varepsilon \Psi , 
\end{align}
where $\Psi = \left(u_{\uparrow}, u_\downarrow ,v_\uparrow,
  v_\downarrow \right)^{\rm T}$ is a 4-component wave function with
separate amplitudes for the particle/hole excitations ($u$, $v$) in
the spin up/down bands ($\uparrow$, $\downarrow$). The Bogoliubov-de
Gennes Hamiltonian $\mathcal H$ has the general form
\begin{align}
  \mathcal{H}&=\left(\begin{array}{cc}
      \hat H_0 & i\sigma^{2}\Delta(\vr)\\
      -i\sigma^{2}\Delta(\vr)^* &
      - \hat H_0^{\ast}\end{array}\right).
\label{eq:bdg-hamiltonian}
\end{align}
Here, the superconducting order parameter $\Delta(\vr) = \Delta_0
e^{i\phi} \Theta(z)$, where $\Theta(z) = 1$ if $z > 0$ and $0$
otherwise, and the $\sigma_i$ are the Pauli matrices, $i=1,2,3$. We
take the normal-state Hamiltonian $H_0$ to be of the form
\begin{eqnarray}
  \hat H_0 = 
  \frac{\vp^2}{2 m} - \sum_{\sigma=\up,\down} \mu_{\sigma}(z) \hat P_{\sigma} 
  + \hbar w \delta(z) + \hat H_{\rm SO},
\end{eqnarray}
where $m$ is the electron mass (taken to be the same in H and S),
\begin{equation}
  \pot_{\sigma}(z) = \left\{ \begin{array}{ll}
  \pot_{{\rm H}\sigma} & \mbox{if $z < 0$}, \\
  \pot_{{\rm S}} & \mbox{if $z > 0$}, \end{array} \right.
\end{equation}
with the potentials $\pot_{{\rm H}\uparrow}$, $\pot_{{\rm
    H}\downarrow}$, and $\pot_{\rm S}$ representing the combined
effect of the chemical potential and band offsets for the majority and
minority electrons in the half metal and for the superconductor,
respectively, and where $w$ sets the strength of a delta-function
potential barrier at the interface. The operators
\begin{equation}
  \hat P_{\uparrow} = \frac{1}{2}(1 + \sigma_3), \ \
  \hat P_{\downarrow} = \frac{1}{2}(1 - \sigma_3)
\end{equation}
project onto the majority and minority components, respectively. (The
magnetization direction in H is taken as the spin quantization axis,
which need not coincide with the z-axis.)  We will take the limit
$\pot_{{\rm H}\downarrow} \to -\infty$, such that only the majority
spin band is present in H. We further write
\begin{equation}
  \pot_{{\rm H}\up} = \frac{\hbar^2 k_{\rm F,H}^2}{2m},\ \
  \pot_{{\rm S}} = \frac{\hbar^2 k_{\rm F,S}^2}{2m},
\end{equation}
where $k_{\rm F,H}$ and $k_{\rm F,S}$ are the Fermi wavenumbers in H and S, respectively, and
\begin{equation}
  \pot_{{\rm H}\downarrow} = -\frac{\hbar^2 \kappa^2}{2 m},
\end{equation}
where $\kappa$ is the minority wavefunction decay rate. The Fermi velocities are defined as
\begin{equation}
  v_{\rm F,H} = \hbar k_{\rm F,H}/m,\ \
  v_{\rm S,H} = \hbar k_{\rm S,H}/m.
\end{equation}
The step function model for the superconducting order parameter
$\Delta(\vr)$ is justified for $s$-wave superconductors if the
coupling to the half metal takes place via a tunnel barrier with
transparency $\tau \ll 1$,\cite{Likharev1979} which corresponds to the
requirement that $|w| \gg v_{\rm F,H}$, $v_{\rm F,S}$.

The operator $\hat H_{\rm SO}$ represents the effect of spin-orbit
coupling. We consider the case that $\hat{H}_{\rm SO}$ is linear in the
momentum $\vp$ and that $\hat H_{\rm SO}$ is nonzero in S
only,\cite{Note1}
\begin{equation}
  \hat H_{\rm SO} =
  \frac{\hbar}{2}[\vp \Theta(z) + \Theta(z) \vp] \cdot 
  \sum_{i=1}^{3} \vOmega_i
  \sigma_i ,
\end{equation}
where we denote $\boldsymbol{\Omega}_i = (\Omega_{i,x}, \Omega_{i,y},
\Omega_{i,z})^T$. Such a SOI may originate from the breaking of
inversion symmetry by the crystal structure of S or due to an
inversion asymmetry of the HS heterosystem. We assume that the
spin-orbit interaction is weak, $\hbar |\boldsymbol{\Omega}(\mathbf
k_{\rm F,S})| \ll v_{\rm F, H}$, $v_{\rm S,H}$, so that it can be captured
by treating $\hat H_{\rm SO}$ to first order in perturbation theory.

In addition to the spin-singlet order parameter contained in Eq.\
(\ref{eq:bdg-hamiltonian}), the presence of SOI generally allows for a
triplet contribution to the order parameter, which is of the form
$\Delta(\vp) = \sum_{i=1}^{3} \Delta_i(\vp) \sigma_i$. Because of the
Pauli principle, these triplet components are odd in momentum,
$\Delta_i(-\vp) = - \Delta_i(\vp)$. Such a triplet contribution is
absent if the pairing interaction is isotropic,\cite{Tinkham2004} but
it may be present if the pairing interaction is
anisotropic.\cite{Gorkov2001} In appendix~\ref{sec:triplet-pairings-s}
we include triplet pairings in the model and give the results for the
Andreev reflection amplitudes. The spin-orbit interaction does not
lead to a modification of the magnitude of the spin-singlet order
parameter to first order in $\hat H_{\rm SO}$.

\subsection{Andreev reflection amplitudes}
\label{sec:s-matrix-hm}

We now calculate the Andreev reflection amplitudes for of the
interface between \HM and S using the Blonder-Tinkham-Klapwijk
formalism.\cite{Blonder1982} At the HS interface triplet Andreev
reflection occurs because quasiparticles incident on the interface
from \HM penetrate the superconductor over a finite length before
being reflected. Due to the SOI, spin is not a good quantum number in
S, which makes spin-flip reflection possible. The Andreev reflection
amplitudes are found by matching eigenfunctions of $\mathcal H$ in \HM
and S to linear order in the SOI. (An alternative method, using
perturbation theory in the SOI Hamiltonian, will be described at the
end of this subsection.) The matching conditions, continuity and
conservation of particle flux, hold for plane-wave eigenstates in the
immediate proximity of the interface on length scales of the Fermi
wavelength. Thus, the S-matrix of the interface is a local property
and will not be changed by weak disorder.

Starting point of the matching procedure are expressions for the
general solutions of the Bogoliubov-de Gennes equation in \HM and S,
near the HS interface. Because of translation symmetry along the
interface, we can consider plane-wave solutions with wavenumbers $k_x$
and $k_y$ in the $x$ and $y$ directions parallel to the interface. In
H, one then finds six linearly independent solutions, which we label
$\Psi_{{\rm e},\uparrow,\pm}$, $\Psi_{{\rm h},\uparrow,\pm}$, $\Psi_{{\rm
    e},\downarrow}$, and $\Psi_{{\rm h},\downarrow}$,
\begin{align}
  \Psi_{{\rm e},\uparrow,\pm}(\vr) &=
  \frac{e^{\pm i k_{z}(+\varepsilon) z + i k_x x + i k_y y}}{\sqrt{v_{z}(\varepsilon)}}
\left(1,0,0,0 \right)^{\rm T}
, 
  \label{eq:PsiH1} \\
  \Psi_{{\rm h},\uparrow,\pm}(\vr) &=
  \frac{  e^{\mp i k_{z}(-\varepsilon) z + i k_x x + i k_y y}}{\sqrt{v_{z}(-\varepsilon)}}
  \left(0, 0,  1,  0\right)^{\rm T}
,~~~ \\
  \Psi_{{\rm e},\downarrow}(\vr) &=
  e^{\kappa_{z}(+\varepsilon) z + i k_x x + i k_y y}   \left(0, 1,  0,  0\right)^{\rm T}, \\  
  \Psi_{{\rm h},\downarrow}(\vr) &=
  e^{\kappa_{z}(-\varepsilon) z + i k_x x + i k_y y}
  \left(0, 0,  0,  1\right)^{\rm T}, 
  \label{eq:PsiH4}
\end{align}
where $k_{z}(\varepsilon)$ and $\kappa_z(\varepsilon)$ are the
positive solutions of
\begin{eqnarray}
  k_{z}(\varepsilon)^2 &=& {k_{\rm F,H}^2
  - k_{||}^2 + 2 m \varepsilon/\hbar^{2}}, \nonumber \\
  \kappa_{z}(\varepsilon)^2 &=& {\kappa^2
  +k_{||}^2 - 2 m \varepsilon/\hbar^{2}}, 
  \label{eq:kz}
\end{eqnarray}
and
\begin{equation}
  \vkp = (k_x,k_y,0)^{\rm T},
\end{equation}
is the momentum parallel to the interface and
\begin{equation}
  v_{z}(\varepsilon) = {\hbar k_{z}(\varepsilon)}/{m}.
\end{equation}
The states labeled with $+$ and $-$ are majority states moving towards
or away from the interface, respectively. They are normalized to unit
flux. The states labeled with $\downarrow$ are minority states that
decay into the half metal. They appear in intermediate stages of the
calculation only and their normalization is not important.

Only the spin-orbit interaction terms proportional to $\vOmega_1$ and
$\vOmega_2$ give rise to spin flips in the superconductor. For a
calculation of Andreev reflection amplitudes linear in the SOI, it is
then sufficient to set $\vOmega_3=0$, which significantly simplifies
the form of the solutions of the Bogoliubov-de Gennes equation in
S. In S, one then finds four linearly independent solutions
$\Psi_{s,t}$ of the Bogoliubov-de Gennes equation, $t,s = \pm 1$,
which read
\begin{eqnarray}
  \Psi_{s,t}(\vr) &=&
  \frac{1}{2}
  \left( \begin{array}{c}
  1 \\ e^{i \gamma_{s,t}} \\ e^{-i \phi - i s \eta + i \gamma_{s,t}} \\
  e^{-i \phi - i s \eta} \end{array} \right)
  e^{i q_{s,t} z + i k_x x + i k_y y},~~~
  \label{eq:PsiS1}
\end{eqnarray}
where
\begin{eqnarray}
  \eta &=& \arccos(\varepsilon/\Delta_0), \\
  q_{s,t} &=& t \sqrt{k_{{\rm F,S}}^2 - k_{||}^2 + 2 i t m
    \sqrt{\Delta_0^2-\varepsilon^2}} - m s t
  \Omega_{s,t} \,  , \quad \mbox{} {}
\end{eqnarray}
and $\gamma_{s,t}$ and $\Omega_{s,t}$ are defined such that
\begin{equation}
  (\vOmega_1 + i \vOmega_2) \cdot \vq =
  \Omega_{s,t} q e^{i \gamma_{s,t}},
\end{equation}
with $\vq = (k_x,k_y,q_{s,t})^{\rm T}$.

A complete solution $\Psi(\vr)$ of the Bogoliubov-de Gennes equation
consists of a linear combination of the six special solutions
(\ref{eq:PsiH1})--(\ref{eq:PsiH4}) in H for $z < 0$ and a linear
combination of the four special solutions (\ref{eq:PsiS1}) in S, with
the boundary conditions
\begin{eqnarray}
\label{eq:bc-cont}
  \Psi(\vr) \Big|_{z=0^-} &=& \Psi(\vr)
  \Big|_{z = 0^+} \\
  \frac{\partial \Psi(\vr)}{\partial z}
  \Big|_{z =0^-}  &=&
  \frac{\partial \Psi(\vr)}{\partial z}
  \Big|_{z = 0^+} \nonumber \\ && \mbox{}
  + m  \left(i \sum_{j=1}^{3} \Omega_{j,z} \sigma_j
  + \frac{2 w}{\hbar} \right) \Psi(\vr)\Big|_{z = 0^+} \nonumber \\
\label{eq:bc-veloc}
\end{eqnarray}
at the interface $z=0$. Since $\Psi(\vr)$ is a four-component spinor,
the boundary conditions provide eight linear relations between the
coefficients of the ten basis functions. The Andreev amplitude
$r_{\eh}$ is then defined as the coefficient of $\Psi_{{\rm
    h},\uparrow,-}$ if the coefficients of the two incoming wave
solutions $\Psi_{{\rm e},\uparrow,+}$ and $\Psi_{{\rm h},\uparrow,+}$
are chosen to be $1$ and $0$, respectively. Analogously, the Andreev
amplitude $r_{\he}$ is defined as the coefficient of $\Psi_{{\rm
    e},\uparrow, -}$ if the coefficients of $\Psi_{{\rm
    e},\uparrow,+}$ and $\Psi_{{\rm h},\uparrow,+}$ are chosen to be $0$ and
$1$, respectively.  To lowest order in the SOI and to lowest order in
the normal-state transmission $\tau(\theta)$ of the HS interface, we
find that
\begin{align}
  \label{eq:reh}
  r_{\eh}(\vkp,\varepsilon) = &
  \frac{-im\tau(\theta)e^{- i \phi }
    \left(\vOmega_{1}+i\vOmega_{2}\right) \cdot \vkp \Delta_{0}}
  {2(k_{\rm F,S}^{2}- k_{\rm F,H}^{2}\sin^{2} \theta)
  \sqrt{\Delta_{0}^{2}-\varepsilon^{2}}}
\end{align}
and
\begin{eqnarray}
  \label{eq:rhe}
  r_{\he}(\vkp,\varepsilon) &=& r_{\eh}(-\vkp,-\varepsilon)^\ast \nonumber \\ &=&
  \frac{-im\tau(\theta)e^{i \phi }
    \left(\vOmega_{1}-i\vOmega_{2}\right) \cdot \vkp \Delta_{0}}
  {2(k_{\rm F,S}^{2}- k_{\rm F,H}^{2}\sin^{2} \theta)
  \sqrt{\Delta_{0}^{2}-\varepsilon^{2}}}  
\end{eqnarray}
Here $\theta = \arcsin(|\vkp|/k_{\rm F,H})$ is the angle between the
incident momentum and the interface normal, see
Fig.~\ref{fig:HM-S-interface}, and  
\begin{eqnarray}
  \label{eq:transmission-amplitude}
  \tau(\theta)^2 
  &=& \frac{v_{\rm F,H}^2 \cos^2 \theta
    (v_{\rm F,S}^{2}- v_{\rm F,H}^{2} \sin^{2}\theta)}
    {w^{4}} + {\cal O}\left(\frac{1}{w^{6}} \right) . \quad \mbox{} \,{}
\end{eqnarray}
Equation (\ref{eq:reh}) has been simplified using the ``Andreev
approximation'', which amounts to neglecting corrections of order
$\mathcal O(\varepsilon/E_{\rm F,S}, \Delta_0/E_{\rm F,S})$. (This
approximation is uniformly valid for all angles if $k_{\rm F,S} >
k_{\rm F,H}$. If $k_{\rm F,S} \le k_{\rm F,H}$ there is a small range
of angles for which the approximation fails.)

The divergence for $\varepsilon \to \pm \Delta$ in Eqs.\
(\ref{eq:reh}) and (\ref{eq:rhe}) is a consequence of the expansion in
the normal-state transmission coefficient $\tau$ of the HS interface
and has to be cut off for $1 - (\varepsilon/\Delta)^2 \lesssim
\tau^2$. This means that the immediate vicinity of $\pm \Delta$ has to
be excluded from the region of validity of Eqs.\ (\ref{eq:reh}) and
(\ref{eq:rhe}), so that these equations are valid for $1 -
(\varepsilon/\Delta)^2 \gg \tau^2$ only. The same condition will be
required for the validity of Eq.\ (\ref{eq:ree}) below and for
expressions that are derived from these equations. (We note that
similar restrictions also apply to an expansion in the transmission
coefficient for a normal-metal--superconductor interface, see, {\em
  e.g.}, Ref.\ \onlinecite{Blonder1982}.)

For completeness, we also give the results for the normal reflection
amplitudes $r_{\rm ee}$ and $r_{\rm hh}$ of the HS interface
consistent with the assumptions of our calculation
\begin{align}
  \label{eq:ree}
  &r_{\rm ee}(\vkp,\varepsilon) = -1 + i  \sqrt{\frac{\tau(\theta) k_{\rm F,H}
      \cos\theta}{ \sqrt{k_{\rm F,S}^2 - k_{\rm F,H}^2 \sin^2
    \theta} } } \\ \notag
& \quad \quad + \frac{\tau(\theta)}{2} \left(\frac{ k_{\rm F,H}
      \cos\theta}{\sqrt{k_{\rm F,S}^2 - k_{\rm F,H}^2 \sin^2
    \theta }} - \frac{i \varepsilon}{\sqrt{\Delta_0^2 -
      \varepsilon^2}}\right) \, , \\ 
\label{eq:rhh}
& r_{\rm hh}(\vkp,\varepsilon) = r_{\rm
  ee}^\ast (-\vkp, -\varepsilon).
\end{align}

Alternatively, Eqs.~(\ref{eq:reh}) and (\ref{eq:rhe}) can be obtained
from a calculation of the first-order perturbation theory correction
to the scattering matrix of the HS interface without spin-orbit
interaction. This calculation is outlined in the appendix. (See
Ref.~\onlinecite{Kupferschmidt2010} for more details.)

Equations (\ref{eq:reh}) and (\ref{eq:rhe}) are the two central
results of the first part of this article. Although the Andreev
reflection amplitudes have been derived for a specific model
Hamiltonian and in the limit of a tunneling interface, we believe that
the symmetry properties of $r_{\eh}$ and $r_{\he}$ --- $r_{\eh}$ and
$r_{\he}$ are {\em odd} in $\vkp$ and {\em even} in $\varepsilon$ ---
persist in a more general calculation, as long as the SOI is linear in
momentum. We have verified this statement for the cases that a finite
minority-wavefunction decay rate $\kappa$ is included in the
calculation, that the spin-orbit interaction extends only a finite
distance into the superconductor, and that higher-order terms in the
interface transmission $\tau(\theta)$ are included. (See
App. \ref{sec:eigenstates-hm} for details.)

The antisymmetry of $r_{\eh}$ and $r_{\he}$ as a function of $\vkp$
implies that the Andreev reflection amplitudes $r_{\eh}$ and $r_{\he}$
contain only four elements $\Omega_{1,x}$, $\Omega_{1,y}$,
$\Omega_{2,x}$, and $\Omega_{2,y}$ of the spin-orbit coupling matrix.
We had already discussed, that the three elements $\Omega_{3,x}$,
$\Omega_{3,y}$, and $\Omega_{3,z}$ that describe the coupling between
the spin component parallel to the magnetization direction and the
orbital motion of the electrons do not give rise to spin flips and,
hence, do not contribute to the Andreev reflection
amplitude. Equations (\ref{eq:reh}) and (\ref{eq:rhe}) show that the
same is true for the elements $\Omega_{1,z}$, $\Omega_{2,z}$, and
$\Omega_{3,z}$ of the spin-orbit coupling matrix that couple the
electron spin to the orbital motion perpendicular to the interface
and, thus, provide a spin-flip mechanism that is symmetric in
$\vkp$. For zero excitation energy $\varepsilon$, this observation can
be understood from the general symmetry considerations of Ref.\
\onlinecite{Kupferschmidt2010}, which state that $r_{\rm he}(\vkp,0) =
0$ if $r_{\rm he}$ is a symmetric function of $\vkp$. That this
remains true for nonzero $\varepsilon$ is special to the case of
spin-orbit coupling as a source of spin-flip scattering and requires
the explicit calculation of this section.

There is a direct relation between the Andreev reflection amplitudes
$r_{\eh}$ and $r_{\he}$ and the anomalous Green function $f(\vk,i
\omega)$,\cite{Kupferschmidt2010}
\begin{equation}
  f(\vk,i\omega) \propto 
  \left\{ \begin{array}{ll}
  \Theta(-k_z) r_{\he}(\vkp,i \omega) & \mbox{if $\omega > 0$}, \\
  -\Theta(k_z) r_{\eh}(\vkp,-i \omega)^* & \mbox{if $\omega < 0$},
  \end{array} \right.
\end{equation}
up to a prefactor that is not important for the identification of the
symmetries of $f$.  Since $r_{\eh}(\vkp,-i \omega)^* = -r_{\he}(\vkp,i
\omega)$ in the present case, see Eq.\ (\ref{eq:rhe}), one concludes
that the induced superconducting correlations in H are odd in momentum
({\em i.e.}, predominantly of (complex) $p$-wave type) and even in
frequency.

\section{Applications: Subgap conductance and Josephson
  current}\label{sec:subg-cond-josephs}

As an application, we now calculate the subgap conductance of an HS
junction and the Josephson current of a
superconductor--half-metal--superconductor (SHS) junction.

\subsection{Subgap conductance}

We assume the interface to have lateral dimensions $W_x \times W_y$
and impose periodic boundary conditions in these directions. This
leads to a quantization of the transverse modes with wave numbers
$k_{n_x} = 2 \pi n_x/W_x$ and $k_{n_y} = 2 \pi n_y/W_y$, $n_x$ and
$n_y$ integer. At zero temperature, the differential conductance $G =
dI/dV$ can be calculated in terms of the Andreev reflection amplitudes
$r_{\eh}$. Replacing the summation over modes by an integral, we
find\cite{Blonder1982,Takane1992}
\begin{align}
  \label{eq:G-serial}
  G(V) &=  \frac{2 e^2}{h} 
 \mathrm{Tr}
  |r_{\eh}(\vkp,e V)|^2 \, ,
\end{align}
where $\mathrm{Tr} \{\ldots\}= W_x W_y/(4 \pi^2) \int_{k_{\parallel} <
  k_{\rm F,H}} d\vkp \{\ldots\} $ is the trace over transverse modes
$\vkp$.  The factor two is due to the doubling of the transferred
charge by conversion of an electron into a hole upon Andreev
reflection.

Substituting Eq.\ (\ref{eq:reh}) for the Andreev reflection amplitude
$r_{\eh}$ and performing the integrations over $k_x$ and $k_y$, we
then find
\begin{eqnarray}
  \label{eq:G-serial-res}
  G(V) &=& \mbox{}  \frac{e^2}{h} \frac{3 N }{8} \frac{\left\langle
      \tau^2\right\rangle_\theta \Delta_0^2}{\Delta_0^2
    -(eV)^2}
  \nonumber \\ && \mbox{} \times
   \frac{\hbar^2 (\Omega_{1,x}^2 + \Omega_{2,x}^2 + \Omega_{1,y}^2 + \Omega_{2,y}^2) }{v_{\rm F,H}^2} 
\end{eqnarray}
where $N =k_{\rm F,H}^2 W_x W_y/4\pi$ is the number of propagating
channels in H and
\begin{equation}
  \langle \ldots \rangle_\theta = 
   2 \int_{0}^{\pi/2} d \theta \ldots \sin \theta \cos \theta
\end{equation}
denotes an average over the angle of incidence $\theta$. 

\subsection{Josephson current}

\begin{figure}
  \centering
  \flushleft a \\
  \begin{minipage}[c]{1.0\linewidth}
    \includegraphics[width = 0.85 \linewidth]{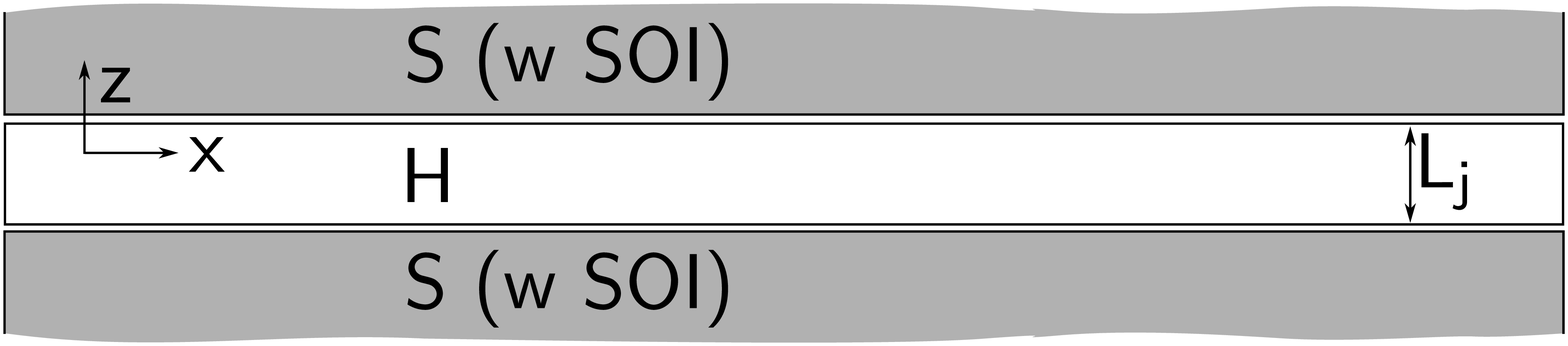}\\
  \end{minipage}
  
  \bigskip
  b \\
  \begin{minipage}[b]{1.0\linewidth}
    \includegraphics[width = 0.85 \linewidth]{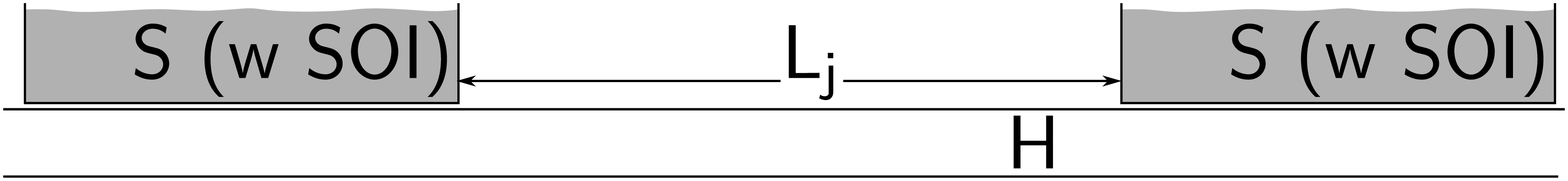}
  \end{minipage}
  \caption{Serial (top, a) and a lateral (bottom, b) superconductor--half-metal--superconductor (SHS) junction. }
  \label{fig:josephson-serial-lateral}
\end{figure}

For the Josephson effect, we consider a half-metallic junction of
length $L_{\rm j}$ separating two superconducting contacts. Again, we
take the junction to have lateral dimensions $W_x \times W_y$ and
impose periodic boundary conditions in the $x$ and $y$
directions. Taking periodic boundary conditions is justified if the
lateral dimensions $W_{x,y} \gg L_{\rm j}$, see Fig.\
\ref{fig:josephson-serial-lateral} a. We further take both HS
interfaces to have the same normal-state transmission $\tau(\theta)$,
take the same spin-orbit interaction Hamiltonian $\hat H_{\rm SO}$ in
both superconductors, and neglect impurity scattering in the
half-metallic junction.

The Josephson current can be found from the density of states in the
junction which, in turn, may be expressed in terms of the scattering
matrices of the HS interfaces (see
Ref.~\onlinecite{Beenakker1991,Brouwer1997} for details). In the
absence of impurity scattering, the contributions to the Josephson
current from different transverse wavevectors $\vkp$ add up, and one
finds that the Josephson current at temperature $T$ is given by the
expression
\begin{eqnarray}
  \label{eq:josephson-general}
  I &=& - \frac{2 e k_{\rm B} T}{ \hbar}
  \frac{d}{d \phi} 
  \sum_{n=0}^\infty  
  \nonumber \\ &&  \times
  \mathrm{Tr}\Big\{\ln \det 
  \left[ 1 - \mathcal R(\vkp,i \omega_n) 
  \mathcal R'(\vkp,i \omega_n) \right] \Big\},
\end{eqnarray}
where $\omega_n = (2n + 1) \pi k_{\rm B} T$ is the Matsubara frequency,
$\mathcal R$ is a $2 \times 2$ reflection matrix containing reflection
and transmission amplitudes for the first HS interface,
\begin{eqnarray}
  \mathcal R(\vkp,i \omega_n) &=&
  \left( \begin{array}{cc}
  e^{i k_{z}(i \omega_n) L_{\rm j}} & 0 \\
  0 & e^{-i k_{z}(-i \omega_n) L_{\rm j}}
  \end{array} \right)
  \nonumber \\ && \mbox{} \times
  \left( \begin{array}{cc}
  r_{\rm ee}(\vkp,i \omega_n) &
  r_{\rm eh}(\vkp,i \omega_n) \\
  r_{\rm he}(\vkp,i \omega_n) &
  r_{\rm hh}(\vkp,i \omega_n) \end{array} \right),~~~
\end{eqnarray}
with $k_z(\varepsilon)$ is given in Eq.\ (\ref{eq:kz}) and the
reflection amplitudes given by Eqs.\ (\ref{eq:reh})--(\ref{eq:rhh})
above, and ${\cal R}'(\vkp,i \omega_n)$ is a similar matrix describing
Andreev reflection at the second HS interface. Specifically, in the
serial geometry in Fig.~\ref{fig:josephson-serial-lateral} a, ${\cal
  R}'(\vkp,i \omega_n) = {\cal R}(\vkp,i \omega_n)|_{\phi=0}$ with the
phase of the order parameter set to zero.

Closed-form expressions for the Josephson current $I$ can be obtained
in limiting cases only. In the limit of ``long'' junction, $L_{\rm j}
\gg \xi$, where $\xi = \hbar v_{{\rm F,H}}/\Delta_0$ is the
superconducting coherence length, and for high temperatures $k_{\rm B} T
L_{\rm j}/\hbar v_{\rm F,H} \gg 1$ one finds
\begin{align}
  \label{eq:josephson-serial-high-t}
  e I =&\frac{2 e^2}{h}   8 \pi k_{\rm B} T \sin(\phi) 
  \\  & \times  \mathrm{Tr} \left\{  |r_{\eh} (\vkp,
  i\omega_0)|^2_{\phi=0} \, e^{2 \omega_0 L/v_z(0)} \right\} \notag
\end{align}
Performing the integration over the transverse momentum $\vkp$ for the
parabolic dispersion of our model Hamiltonian then gives the result
\begin{align}
  e I = \frac{2  \pi k_{\rm B} T \sin \phi}{3} G(0) f(2 \pi k_{\rm B} T
  L_{\rm j}/\hbar v_{\rm F,H}),
\end{align}
where $G(0)$ is the zero-bias conductance of a single HS interface,
see Eq.\ (\ref{eq:G-serial-res}), and
\begin{align}
  f(x) &= e^{-x} (6 - 10 x-x^2+x^3) + x^2 (x^2 - 12)
\mathrm{Ei} (-x)\notag  \\ & \approx  48 e^{-x}/x^2 \quad \mathrm{for} \,
x\gg 1, 
\end{align}
with the exponential integral $\mathrm{Ei}(x) = -\int^\infty_{-x} dt e^{-t}/t $. 

In the opposite limit of zero temperature, the expression for the
Josephson current in a long junction ($L_{\rm j} \gg \xi$) becomes
\begin{align}
  \label{eq:josephson-serial-zero-t}
  I  & = \frac{2 e \sin \phi}{\hbar }
  \int_{0}^{\infty} 
  d\omega \,    \\ &\times \mathrm{Tr}\Bigg\{  \frac{
    |r_{\eh}|^2  }{\cosh \frac{2 \omega L_{\rm j}}{v_z}  -
    \mathrm{Re} \left[ r_{\rm ee}^2 e^{2 i k_z(0) L }  \right]+ |r_{\eh}|^2
    \cos \phi} \Bigg\} \notag \, .
\end{align}
Normal reflection with amplitude $r_{\rm ee}$ at the two HS interfaces
gives rise to terms in Eq.~(\ref{eq:josephson-serial-zero-t}) that
oscillate with the junction length $L_{\rm j}$. These oscillations
disappear once the trace over transverse modes is taken, since $k_z(0)
L_{\rm j} \gg 1$ for long junctions. The remaining non-oscillatory
contribution to the supercurrent $\bar I$ can then be calculated by
taking the average $\bar I = (2 \pi)^{-1} \int_0^{2 \pi} d\chi
I(\chi)$, where $I(\chi)$ is obtained from Eq.\
(\ref{eq:josephson-serial-zero-t}) by the replacement $2 i k_z(0)
L_{\rm j} \to
\chi$. 
One thus obtains
\begin{align}
  \label{eq:josephson-serial-zero-t-2}
   I  & = \frac{e \sin \phi}{2 L_{\rm j}}   \mathrm{Tr} \left\{ 
      v_z |r_{\eh}|^2 
   \log\left[ 
    \frac{16 \sin^2(\phi/2)}{|r_{\eh}|^2 \sin^2 \phi }\right] \right\} .
\end{align}
The remaining trace over modes can be performed to logarithmic
accuracy by neglecting the dependence of the argument of the logarithm
on $\vkp$. This amounts to the replacement $|r_{\eh}|^2 \to \langle
|r_{\eh}|^2 \rangle_{\theta} = h G(0)/2 e^2 N$ in the
argument of the logarithm. One then obtains
\begin{align}
  \label{eq:josephson-serial-zero-t-3}
  eI & = \frac{4\pi \hbar v_{\rm F}}{15 L_{\rm j}} G(0)
  \sin \phi \log
  \left[ \frac{32 e^2 N \sin^2(\phi/2)}{h G(0)
      \sin^2 \phi}\right],
\end{align}
up to corrections of order $G(0) \hbar v_{\rm F}/L_{\rm j}$, but
without the large logarithm $\log (e^2 N/G(0) h))$.  The small
corrections to the approximately sinusoidal phase dependence of the
supercurrent in Eq.~(\ref{eq:josephson-serial-zero-t-3}) originate
from scattering processes with multiple normal reflections at the HS
interfaces.

\section{Lateral geometry}\label{sec:lateral-geometry}

An experimentally relevant situation\cite{Keizer2006,Anwar2010} is the
lateral geometry where the superconducting contact is attached
laterally to a thin \HM film. This situation is shown in
Fig.~\ref{fig:HM-S-interface}b. In comparison to the serial contact
considered in the previous section, a lateral contact 
has a significantly larger contact area per unit cross section of
H. Multiple reflections occur at the \HM-S interface, because
quasiparticles are repeatedly reflected backwards from the lower film
boundary towards the interface. In the absence of impurity scattering
in the half-metallic film, the coherent addition of these multiple
Andreev reflections leads to a significant enhancement of the Andreev
reflection probability for a quasiparticle incident on the lateral
contact from the left (in Fig.\ \ref{fig:HM-S-interface}b), as we now
show.

We choose coordinates, such that the half metal occupies the region
between $z=0$ and $z=-d$ and the superconductor occupies the region $x
> 0$, $z > 0$, see Fig.\ \ref{fig:HM-S-interface}b. We take periodic
boundary conditions in the $y$ direction, with system size $W_y$. For
the half metal, we take hard-wall (Dirichlet) boundary conditions at
$z=-d$ for all $x$, and at $z=0$ for $x < 0$. The thickness $d$ of the
half-metallic film is taken to be much smaller than the
superconducting coherence length $\xi = \hbar v_{{\rm
    F,H}}/\Delta_0$. As before, we take the HS interface to be a
tunneling interface with a transmission probability $\tau(\theta) \ll
1$.

The goal of our calculation is to find the amplitude $r_{\rm
  he}^{\lat}$ that a right-moving electron-like quasiparticle
approaching the contact from the left is Andreev reflected into a
left-moving hole-like quasiparticle, as well as the amplitude $r_{\rm
  eh}^{\lat}$ for the process that a hole-like quasiparticle is
Andreev reflected as an electron-like quasiparticle. The calculation
proceeds in three steps: First, we construct scattering states in the
absence of spin-orbit interaction; Second, we account for the effect
of spin-orbit interaction in a superconducting region of length $d \ll
\delta L \ll \hbar v_{{\rm F,H}}/\Delta_0$ using perturbation theory;
Finally, we combine Andreev reflections from different segments and
compute the Andreev reflection amplitudes $r_{\rm he}^{\lat}$ and
$r_{\rm eh}^{\lat}$.

\subsection{Scattering states in the absence of SOI}

Because of translation symmetry in the $y$ direction, the scattering
states can be chosen as plane waves in the $y$ direction with
wavenumber $k_y$, which takes discrete values only because of the
periodic boundary conditions in the $y$ direction. We first construct
scattering states for $x < 0$. There, because of the hard-wall
boundary conditions at $z=0$ and $z=-d$, the $z$-dependence can be
chosen proportional to $\sin(k_z z)$, where $k_z = n \pi/d$,
$n=1,2,\ldots$, is discrete, too. For each discrete value of the
transverse momenta $\vkprp = (0,k_y,k_z)^{\rm T}$ one then has four
scattering states which we label $\Phi_{{\rm e},\vkprp\pm}$ and
$\Phi_{{\rm h},\vkprp,\pm}$,
\begin{eqnarray}
  \Phi_{{\rm e},\vkprp,\pm}(\vr) &=&
  \frac{2   e^{\pm i k_x(\varepsilon) x + i k_y y}
  \sin (k_z z)}{\sqrt{v_{x} d W_y}}
  \left(\begin{array}{c}
  1\\
  0\\
  0\\
  0\end{array}\right)
  , \nonumber \\
  \label{eq:Phie}
  \\
  \Phi_{{\rm h},\vkprp,\pm}(\vr) &=&
  \frac{2   e^{\mp i k_{x}(-\varepsilon) x + i k_y y}
  \sin (k_z z)}{\sqrt{v_{x} d W_y}}
  \left(\begin{array}{c}
  0\\
  0\\
  1\\
  0\end{array}\right), \nonumber \\  \label{eq:Phih}
\end{eqnarray} 
where  $k_x(\varepsilon)$ is the positive solution of
\begin{eqnarray}
  k_x(\varepsilon)^2 =
  k_{\rm F,H}^2 - k_y^2 - k_z^2 +
  \frac{2 m \varepsilon}{\hbar^2},
\end{eqnarray}
and 
\begin{equation}
  v_x = \hbar k_x/m.
\end{equation}
The scattering states labeled ``$+$'' represent quasiparticle states
moving to the left; the states labeled ``$-$'' represent quasiparticle
states moving to the right. All scattering states are normalized to
unit flux.

In the region $x > 0$, the scattering states differ from those given
above because of the finite tunnel coupling to the superconductor. In
particular, the scattering states acquire a finite weight inside the
superconductor. In the tunneling limit $\tau \ll 1$, this weight is
small and the majority component of the scattering states inside the
half metal remains well approximated by Eqs.\ (\ref{eq:Phie}) and
(\ref{eq:Phih}) above. The exact expressions for the full scattering
state in the region $x > 0$ are cumbersome, and we refer to Ref.\
\onlinecite{Kupferschmidt2010}, where the detailed expressions can be
found.

The ``turning on'' of the tunnel coupling to the superconductor at
$x=0$ gives rise to a small amount of normal reflection, but it does
not cause Andreev reflection. We neglect this normal reflection at $x
= 0$ in the remainder of this section.

\subsection{Andreev reflection from a superconducting segment of length $\delta L$}

The presence of spin-orbit coupling in the superconductor gives rise
to Andreev reflection at the HS interface, as we have seen in Sec.\
\ref{sec:subg-cond-josephs}. In the second step of our calculation, we
compute the effective Andreev reflection amplitude for a
superconducting segment of size $0 < x < \delta L$. We choose the
length $\delta L$ of the superconducting segment such that $d, \delta
L \ll \xi$. The inequality $d \ll \xi$, together with translation
symmetry in the $y$ direction, ensure that the Andreev reflection
amplitude is diagonal in $k_y$ and $k_z$. The inequality $\delta L \ll
\xi$ gives $|k_x(\varepsilon) - k_x(-\varepsilon)| \delta L \ll 1$ for
excitation energies up to $\Delta_0$. This, in turn, leads to Andreev
reflection amplitudes proportional to $\delta L$, which we write as
$\rho_{\eh}(\vkprp,\varepsilon) \delta L$ and
$\rho_{\he}(\vkprp,\varepsilon) \delta L$, for electron-to-hole and
hole-to-electron reflection, respectively.

Calculating the Andreev amplitudes for the segment $0 < x < \delta L$
in first-order perturbation theory in the spin-orbit interaction gives
\begin{eqnarray}
  \label{eq:rho}
  \rho_{\eh}(\vkprp,\varepsilon) \delta L &=&
  -i \langle \Phi_{{\rm h},\vkprp,-}|\delta \mathcal H_{\rm SO}|
  \Phi_{{\rm e},\vkprp,+} \rangle, \nonumber \\
  \rho_{\he}(\vkprp,\varepsilon) \delta L &=&
  -i \langle \Phi_{{\rm e},\vkprp,-}|\delta \mathcal H_{\rm SO}|
  \Phi_{{\rm h},\vkprp,+} \rangle,
\end{eqnarray}
where $\delta \mathcal H_{\rm SO}$ is the $4 \times 4$ matrix
Hamiltonian representing the projection of the spin-orbit interaction
Hamiltonian onto the segment $0 < x < \delta L$,
\begin{align}
  \delta \mathcal H_{\rm SO} &= \frac{1}{2} \left\{ P_{\delta L}(x) ,
    \mathcal H_{\rm{SO}}
  \right\},
    \label{eq:h-so-bdg-segment} \\
  \mathcal H_{\rm{SO}} &= \left( \begin{array}{cc} \hat H_{\rm SO} & 0
      \\ 0 & - \hat H_{\rm SO}^* \end{array} \right)
  \label{eq:h-so-bdg}
\end{align}
with $P_{\delta L}(x) = 1$ for $0 < x < \delta L$ and $0$ otherwise.
Evaluating the matrix element in the limit $d \ll \delta L \ll \xi$ then gives
\begin{equation}
  \rho_{\eh}(\vkprp,\varepsilon) =
  r_{\eh}(\vkp,\varepsilon) \frac{k_z}{2 d k_{x}},
  \label{eq:deltareh}
\end{equation}
where $\vkp = (k_x(0),k_y,0)^{\rm T}$. Equation (\ref{eq:deltareh})
has the simple interpretation as the Andreev reflection amplitude for
a single reflection at the HS interface, multiplied by the number of
bounces at the HS interface per unit
length.\cite{Kupferschmidt2009,Kupferschmidt2010} Similarly, one finds
that
\begin{equation}
  \rho_{\he}(\vkprp,\varepsilon)  =
  r_{\he}(\vkp',\varepsilon) \frac{k_z}{2 d k_{x}},
\end{equation}
where $\vkp' = (-k_x(0),k_y,0)^{\rm T}$.

In the same way, one also calculates Andreev reflection amplitudes
$\rho_{\eh}(\vkprp,\varepsilon)' \delta L$ and
$\rho_{\eh}(\vkprp,\varepsilon)' \delta L$ for quasiparticles incident
on the segment $0 < x < \delta L$ from the right. These are
\begin{eqnarray}
  \rho_{\eh}(\vkprp,\varepsilon)' &=&
  r_{\eh}(\vkp',\varepsilon) \frac{k_z}{2 d k_{x}}, \\ \rho_{\he}(\vkprp,\varepsilon)' &=&
  r_{\he}(\vkp,\varepsilon) \frac{k_z}{2 d k_{x}}.  
\end{eqnarray}

\begin{figure}
  \centering
  \includegraphics[width = 0.95 \linewidth]{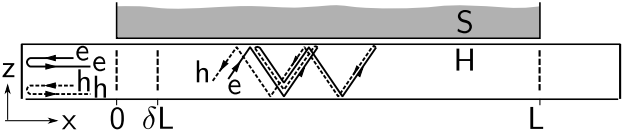}
  \caption{A lateral contact between a half-metallic film or wire and
    a superconductor. The contact has length $L$, with a segment of
    length $\delta L$ singled out. The main text describes how the
    effective Andreev reflection of the entire lateral contact is
    calculated, starting from the effective Andreev reflection
    amplitudes $\rho_{\eh} \delta L$ and $\rho_{\he} \delta L$ of the
    segment of length $\delta L$. As discussed in Sec.\
    \ref{sec:majorana-states}, inserting a normally reflecting
    boundary at the left end of the H film or wire gives rise to a
    Majorana bound state. }
  \label{fig:HM-S-interface-nonlinear}
\end{figure}

\subsection{Effective Andreev reflection amplitudes for the lateral contact}

In the final part of the calculation, we consider a superconducting
contact of finite length $0 < x < L$. In order to keep the notation
simple, we omit the arguments $\vkprp$ and $\varepsilon$ in the
intermediate results.

Upon comparing contacts of length $L$ and $L+\delta L$, one finds,
that
\begin{eqnarray}
  r_{\eh}^{\lat}(L+\delta L) &=&
  r_{\eh}^{\lat}(L)  e^{i (k_x(\varepsilon)-k_x(-\varepsilon)) \delta L} 
  \nonumber \\ && \mbox{}
  + [\rho_{\eh} + \rho_{\he}' r_{\eh}^{\lat}(L)^2] \delta L
  + {\cal O}(\delta L)^2.~~~ 
\end{eqnarray}
Making use of the relations
\begin{equation}
  \rho_{\eh}' = - (\rho_{\he})^*,\ \
  \rho_{\he}' = - (\rho_{\eh})^*,
\end{equation}
which follow from quasiparticle conservation or from the explicit
solutions obtained above, and expanding $k_x(\varepsilon) \approx k(0)
+ \varepsilon/\hbar v_x$ plus terms of order $\varepsilon^2$ that are
neglected in the Andreev approximation, one arrives at a nonlinear
differential equation for $r_{\eh}^{\lat}(L)$,
\begin{eqnarray}
  \frac{d r_{\eh}^{\lat}}{d L} &=&
  \frac{2 i \varepsilon}{\hbar v_x} r_{\eh}^{\lat}
  + \rho_{\eh} - \rho_{\eh}^* (r_{\eh}^{\lat})^2.
\end{eqnarray}
Solving this equation with the boundary condition $r_{\eh}^{\lat}(0)
= 0$ gives
\begin{align}
  r_{\eh}^{\lat}(L) =
  \frac{\rho_{\eh} \sin Q_{\eh} L}{Q_{\eh} \cos Q_{\eh} L - i (\varepsilon/\hbar v_x) \sin
    Q_{\eh} L} 
  \label{eq:rehlat}
\end{align}
and, similarly,
\begin{align}
  r_{\he}^{\lat}(L) =
  \frac{\rho_{\he} \sin Q_{\he} L}{Q_{\he} \cos Q_{\he} L + i
    (\varepsilon/\hbar v_x) \sin Q_{\he} L},
  \label{eq:rhelat}
\end{align}
where we abbreviated
\begin{align}
  Q_{\eh} =& \sqrt{(\varepsilon/\hbar v_x)^2 - |\rho_{\eh}|^2} \nonumber \\
  Q_{\he}=& \sqrt{(\varepsilon/\hbar v_x)^2 - |\rho_{\he}|^2}.
\end{align}

Upon taking the expression for $r_{\eh}^{\lat}(L)$ to first order in
$L$, one verifies that one reproduces the starting point
$r_{\eh}^{\lat}(\delta L) = \rho_{\eh} \delta L$ of the previous subsection. Upon
taking $r_{\eh}^{\lat}(L)$ to first order in $\rho_{\eh}$, one finds
\begin{equation}
  r_{\eh}^{\lat}(L) = \rho_{\eh} \int_0^{L} dx e^{2i \varepsilon x/\hbar v_x},
  \label{eq:rehperturb}
\end{equation}
which one obtains by applying first-order perturbation theory to the
entire superconducting contact of length $L$ at
once.\cite{Kupferschmidt2010} Equation (\ref{eq:rehperturb})
represents the effect of a single Andreev reflection at the HS
interface, with the phase factor accounting for the relative phase
shift between the electron and the Andreev reflected hole for an
Andreev reflection taking place at position $x$. One checks for
consistency that $r_{\eh}^{\mathrm{\lat}}(\delta L) / \delta L$ turns into
Eq.~(\ref{eq:deltareh}) in the limit $\delta L \to 0$.

The relevant limit for the lateral contact of Fig.\
\ref{fig:HM-S-interface}b is the limit $L \to \infty$. In this limit,
the energy dependence of $r_{\rm eh}^{\lat}$ through the energy
dependence of $\rho_{\eh}$ can be neglected in comparison to the
explicit energy dependence in Eq.\ (\ref{eq:rehlat}), so that one may
approximate $\rho_{\eh}(\vkprp,\varepsilon)$ by $\rho_{\eh}(\vkprp,0)$
in Eq.\ (\ref{eq:rehlat}). Defining
\begin{eqnarray}
  \varepsilon_{0}(\vkprp) &=& \hbar v_x |\rho_{\eh}(\vkprp,0)| \nonumber \\ &=&
  \frac{\hbar v_z}{2 d}|r_{\eh}(\vkp,0)|,
\end{eqnarray}
one then finds that
\begin{eqnarray}
  r_{\eh}^{\lat}(\vkprp,\varepsilon) &=& \frac{\rho_{\eh}(\vkprp,0)}{|\rho_{\eh}(\vkprp,0)|}
  e^{i \arcsin(\varepsilon/\varepsilon_0(\vkprp))} \nonumber \\ &=&
  \frac{r_{\eh}(\vkp,0)}{|r_{\eh}(\vkp,0)|}
  e^{i \arcsin \left( \varepsilon/\varepsilon_0(\vkprp) \right)},
  \label{eq:reh-lat-full}
\end{eqnarray}
if $\varepsilon < \varepsilon_0(\vkprp)$.  Hence, as long as
$\varepsilon < \varepsilon_0(\vkprp)$, $|r_{\eh}^{\lat}| =1$ such that
a lateral contact serves as an ``ideal'' contact between a
superconductor and a half metal, allowing perfect spin-flip Andreev
reflection back into the half metal. For $\varepsilon >
\varepsilon_0(\vkprp)$, $r_{\eh}(\vkprp,\varepsilon)$ is an
oscillating function of the contact size $L$, whereas the magnitude of
the Andreev reflection amplitude is a decreasing function of energy,
\begin{eqnarray}
  |r_{\eh}^{\lat}(\vkprp,\varepsilon)|^2 &=&
  \frac{\sin^2 (Q_{\eh} L)}{(\varepsilon/\varepsilon_0(\vkprp))^2 -
    \cos^2 (Q_{\eh} L)}
  \nonumber\\ & \approx &
  1 - \sqrt{1 - (\varepsilon_0(\vkprp)/\varepsilon)^2}
  \ \ \mbox{if $L \to \infty$}, \nonumber \\
  \label{eq:reh-lat-full-high-e}
\end{eqnarray}  
where the last line is obtained by averaging $L$ over a period $0 <
Q_{\eh} L < 2 \pi$ and is proportional to the envelope of
$|r_{\eh}^{\lat}(\varepsilon)|^2$.  The energy $\varepsilon_0(\vkprp)$
separating the regions of complete and partial Andreev reflection can
be interpreted as a mode-dependent proximity-induced ``minigap'' in
the half metallic film.

An analogous calculation including multiple ARs can be done for the
transmission through the contact. For electrons and holes incoming
from the left we find the transmission amplitudes
\begin{align}
  \label{eq:tee-lat-full}
  t_{\rm ee}^{\lat}(\vkprp, \varepsilon) & = 
  \frac{ Q_{\eh} e^{i k_x(0) L}  }{Q_{\eh} \cos Q_{\eh} L - i (\varepsilon/\hbar v_x)
    \sin Q_{\eh} L}  \\
  t_{\rm hh}^{\lat}(\vkprp, \varepsilon) & = 
  \frac{ Q_{\he} e^{-i k_x(0) L}  }{Q_{\he} \cos Q_{\he} L - i (\varepsilon/\hbar v_x)
    \sin Q_{\he} L} 
  \label{eq:thh-lat-full}.
\end{align}
The amplitudes for particles incoming from the right are related to
Eq.~(\ref{eq:tee-lat-full}-~\ref{eq:thh-lat-full}) by $t_{\rm ee}^{\prime
  \lat} = e^{i 2 k_x(0) L} t_{\rm hh}^{\lat}$ and
$t_{\rm hh}^{\prime,\lat} = e^{-i 2 k_x(0) L} t_{\rm ee}^{\lat}$. For a long
contact, $|\rho_{\eh}(\vkprp, \varepsilon)| L \gg 1$, and $\varepsilon <
\varepsilon_0(\vkprp)$ the transmission amplitudes can be approximated as 
\begin{align}
  \label{eq:t-lat-full-asymp}
  t_{\rm ee}^{\lat}(\vkprp, \varepsilon) \approx 2 e^{i k_x(0) L} e^{-q_{\eh} L} e^{i \arcsin(
  \varepsilon/\varepsilon_0(\vkprp) )} , 
\end{align}
with $q_{\eh} = i Q_{\eh}$. Thus, $t_{\rm ee}$, $t_{\rm hh}$ become
exponentially suppressed consistent with a fully developed proximity effect.

\section{Applications: Subgap conductance and Josephson
  current}\label{sec:subg-cond-josephs-1}

The effective Andreev reflection amplitude $r_{\eh}^{\mathrm{\lat}}$
can be used for a calculation of the subgap conductance $G$ and the
Josephson current $I$ in a lateral HS or SHS junction in the same way
as the Andreev reflection amplitude $r_{\eh}$ is used in the case of a
serial junction.

\subsection{Subgap conductance}

In the limit of a long ballistic contact and at the Fermi level
$\varepsilon = 0$, the effective Andreev reflection amplitude of
Eq.~(\ref{eq:reh-lat-full}) has modulus $1$ for all transverse
channels (labeled by the integer $n$ and the wavenumber $k_y$). Hence
the zero bias conductance $G(0)$ is
\begin{align}
  \label{eq:g-lateral-max}
  G(0) = G_{\rm m} = \frac{2e^2}{h} N, 
\end{align}
where
\begin{equation}
  N = \frac{k_{{\rm H,F}}^2 W_y d}{4 \pi}
\end{equation}
is the number of propagating modes at the Fermi level in the half
metal.

Upon increasing $V$, the Andreev reflection probabilities and, hence,
the conductance $G$ decrease. The precise functional form of this
decrease depends on the details of the spin-orbit coupling. For
voltages much larger than the induced ``minigap'' in the half metal,
but still much smaller than $\Delta_0$, i.e., $\hbar v_{{\rm F,H}}
|r_{\eh}|/2 d \ll e V \ll \Delta_0$, we may take
$|r_{\eh}^{\lat}(\vkprp,\varepsilon)|^2$ from
Eq.~(\ref{eq:reh-lat-full-high-e}) and find
\begin{align}
  \label{eq:g-lateral-high-v}
G(V) =& \frac{e^2}{h}     \frac{N
  \langle \tau^2 \rangle_{\theta}}{128} \left( \frac{\hbar}{e V d}\right)^2  \\ \notag
 &\times  \left[
  2\Omega^2_{1,x} + \Omega^2_{1,y}  + 2\Omega^2_{2,x} +
  \Omega^2_{2,y}\right]. 
\end{align}
This decay of the subgap conductance with the applied voltage is a
marked difference with the case of the serial geometry, for which $G$
is an increasing function of $V$.

\subsection{Josephson current}

The expression for the Josephson current $I$ in a lateral SHS junction
can be obtained from Eq.\ (\ref{eq:josephson-general}) upon setting
$r_{\rm ee} = r_{\rm hh} = 0$ and upon replacing $r_{\eh}$ and
$r_{\he}$ by $r_{\eh}^{\lat}$ and $r_{\he}^{\lat}$, respectively,
\begin{eqnarray}
  I &=& - \frac{4 e k_{\rm B} T}{\hbar}
  \frac{W_y d}{4 \pi^2} \mbox{Re}\, \frac{d}{d \phi} 
  \sum_{n=0}^\infty 
  \int_{k_{\perp} < k_{{\rm F,H}}} d\vkprp
  \nonumber \\ && \mbox{} \times
  \ln \left[ 1 - r_{\eh}^{\lat}(\vkprp,i \omega_n)
    r_{\he}^{\lat}(\vkprp,i \omega_n)' e^{-2 \omega_n L_{\rm j}/v_x}
  \right], \nonumber \\ 
  \label{eq:josephson-general2}
\end{eqnarray}
where $L_{\rm j}$ is the junction length, see Fig.\
\ref{fig:josephson-serial-lateral}b,
$r_{\eh}^{\lat}(\vkprp,\varepsilon)$ is the effective electron-to-hole
Andreev amplitude of the right superconducting contact for
quasiparticles incident from the left, and
$r_{\he}^{\lat}(\vkprp,\varepsilon)'$ the effective hole-to-electron
Andreev amplitude of the left superconducting contact for
quasiparticles incident from the right. We have set the phase of the
superconducting order parameter for the right superconductor equal to
zero.
Upon setting $\varepsilon = i \omega_n$, the effective Andreev
reflection amplitudes $r_{\eh}^{\lat}(\vkprp,\varepsilon)$ and
$r_{\he}^{\lat}(\vkprp,\varepsilon)'$ have a well-defined limit for
the contact size $L \to \infty$, which is given by Eq.~(\ref{eq:reh-lat-full}) for all $\omega_n$. 

We first consider the limit when the minigap $\varepsilon_0 \gg E_{\rm
  Th}$ is much larger than the Thouless energy $E_{\rm Th} =\hbar
v_z/L_{\rm j}$ of the junction (long-junction limit). For high temperatures
$k_{\rm B} T L_{\rm j}/\hbar v_{\rm F} \gg 1 $, one finds
\begin{align}
  \label{eq:i-josephson-long-high-T}
  e I = - 8 \sqrt{2 \pi} G_{\rm m} k_{\rm B} T    e^{-\frac{ 2
      \pi k_{\rm B} T L_{\rm j}}{\hbar v_{\rm F,H}}}  \left(\frac{\hbar v_{\rm F,H}}{2
      \pi k_{\rm B} T L_{\rm j}} \right)^{\frac{3}{2}} \sin \phi  \, .
\end{align}
This expression for $I$ has the same sinusoidal phase dependence and
exponential junction length dependence as in the serial geometry, but
in the lateral contact geometry $I$ is proportional to the much
larger conductance $G_{\rm m}$, Eq.~(\ref{eq:g-lateral-max}), instead
of Eq.~(\ref{eq:G-serial-res}). In the limit of zero temperature, the
Josephson current is given by
\begin{align}
  \label{eq:i-josephson-long-long}
  I = \frac{2 e}{3} \frac{N v_{\rm F,H}}{L_{\rm j}}
  \phi, \quad -\pi < \phi < \pi  \, ,
\end{align}
and $I(\phi + 2 \pi) = I(\phi)$.  Equation
(\ref{eq:i-josephson-long-long}) is the known form of a supercurrent
if the superconductor and the normal junction material are strongly
coupled and the junction is disorder-free.\cite{Ishii1970} The phase
dependence is sectionally linear (saw-tooth-like) and the critical
current decreases with the length $L_{\rm j}$ of the junction.

In the opposite limit of a short junction ($\varepsilon_0 \ll E_{\rm
  Th}$), we distinguish three temperature regimes. For very high
temperatures, $k_{\rm B} T \gg E_{\rm Th} = \hbar v_{\rm F,H}/ L_{\rm
  j} $, one obtains a sinusoidal phase dependence of the Josephson
current
\begin{eqnarray}
  \label{eq:j-lat-short-high-t}
  I 
  &=& \frac{3 e N \langle \tau^2 \rangle_\theta}{8 \pi^2 \hbar
    d^2 k_{\rm B} T}\left(\frac{E_{\rm Th}}{2 \pi k_{\rm B} T}\right)^2
  \nonumber \\ && \mbox{} \times \left(
    \Omega^2_{1,x} + \Omega^2_{2,x}\right)  e^{-\frac{2 \pi k_{\rm B} T
    }{E_{\rm Th}}}    \sin \phi.
\end{eqnarray}
For intermediate temperatures, $\varepsilon_0 \ll k_{\rm B} T  \ll E_{\rm Th} $, one finds
\begin{align}
  \label{eq:j-lat-short-medium}
  I =  \frac{e \hbar N \langle \tau^2
    \rangle_\theta}{512 d^2 k_{\rm B} T } \left(
    2 \Omega^2_{1,x} + \Omega^2_{1,y} + 2 \Omega^2_{2,x} + \Omega^2_{2,y} \right) \sin \phi.
\end{align}
For $T= 0$, the trace over transverse modes could not be performed in
closed form. However, the dependence on phase difference $\phi$ can be
found. One obtains
\begin{align}
  \label{eq:j-lat-short-zero-t}
  I = \frac{e}{4  d} \mathrm{Tr}\left\{  v_z |r_{\eh}|
  \right\}\sin \frac{\phi}{2},  \quad -\pi < \phi < \pi  \, ,
\end{align}
and $I(\phi + 2 \pi) = I(\phi)$. The dependence $I \propto \sin
(\phi/2)$ of the zero-temperature Josephson current is reminiscent of
the ``fractional Josephson effect'', characteristic of Josephson
junctions that have Majorana bound states at the superconductor
interfaces\cite{Kitaev2000,kn:fu2009b} (see also the next section).

\section{Majorana states}\label{sec:majorana-states}

Majorana bound states have been proposed as an elementary building
block of a topological quantum computer since they are an example of
an excitation with non-Abelian statistics.\cite{Nayak2008} Majorana
bound states exist as the fundamental excitations of a candidate state
for the $\nu = 5/2$ quantum Hall
effect,\cite{kn:moore1991,kn:read2000} in vortices in superconductors
with a spinless $p$-wave pairing
symmetry,\cite{kn:read2000,kn:ivanov2001,kn:stern2004,kn:stone2006} or
in vortices of $s$-wave superconductors in contact to a topological
insulator\cite{Fu2008} or a standard two-dimensional electron gas in a
large magnetic field and with strong spin-orbit
coupling.\cite{kn:sau2010}

Very recently, it was suggested that Majorana bound states can be
found at the ends of semiconducting quantum wires with strong
spin-orbit coupling and a strong magnetic field, in contact to an
$s$-wave superconductor.\cite{Lutchyn2010,Oreg2010,Alicea2010} In
these proposals, the role of the magnetic field is to create a gap for
spin excitations, so that the wires become effectively half
metallic. We now show that Majorana bound states also occur in the
system considered here: a half-metallic quantum wire in contact to a
superconductor with spin-orbit coupling. This enables us to make
contact between the (experimentally observed) triplet proximity effect
and the so far purely theoretical search for avenues to topological
quantum computation. Our approach has the additional benefit of
providing a fully microscopic description of the $p$-wave proximity
state, in contrast to the existing studies of this effect in
semiconducting wires with strong spin-orbit coupling, which rely on an
effective description using an induced pairing potential in the
semiconducting wire.\cite{Lutchyn2010,Oreg2010}

First, we show that a Majorana state exists at the end of a ballistic
half-infinite half metallic quantum wire laterally coupled to a
superconductor. We consider the geometry shown in
Fig.~\ref{fig:HM-S-interface-nonlinear}. The difference with the
calculation of Sec.\ \ref{sec:subg-cond-josephs-1} is that here the
half metal is a wire, not a film. We therefore have to use hard-wall
boundary conditions in the $y$-direction, not periodic boundary
conditions as in Sec.\ \ref{sec:subg-cond-josephs-1}. With hard-wall
boundary conditions, the Andreev reflection amplitudes $\rho_{\eh}$
and $\rho_{\he}$ per unit length have to be replaced by amplitudes
$\tilde \rho_{\eh}$ and $\tilde \rho_{\he}$, which are defined as
\begin{eqnarray}
  \tilde \rho_{\eh} (k_y, k_z, \varepsilon ) &=& \frac{1}{2} \left[\rho_{\eh} (k_y, k_z,
  \varepsilon ) + \rho_{\eh} (-k_y, k_z, \varepsilon ) \right],~~~ \\
  \tilde \rho_{\he} (k_y, k_z, \varepsilon ) &=& \frac{1}{2} \left[\rho_{\he} (k_y, k_z,
  \varepsilon ) + \rho_{\he} (-k_y, k_z, \varepsilon ) \right].~~~
\end{eqnarray}
Since $\rho_{\eh}$ and $\rho_{\rm he}$ are odd in $k_y$, the
components of the SOI coupling to $k_y$ drop out. Apart from the
replacement $\rho_{\eh} \to \tilde \rho_{\eh}$ and $\rho_{\he} \to
\tilde \rho_{\he}$, the results of Sec.~\ref{sec:lateral-geometry}
continue to hold for the present case.

At the left end of the half-metallic wire, quasiparticles undergo
normal reflection with amplitude $s_{\rm H}(\varepsilon)$ and $s_{\rm
  H}(-\varepsilon)^*$ for electron-like and hole-like quasiparticles,
respectively. With the Andreev reflection amplitudes of
Eqs.~(\ref{eq:reh-lat-full}) we then find a non-degenerate bound state
at $\varepsilon = 0$ with (unnormalized) wavefunction
\begin{align}
\label{eq:majorana-state-inf}
  \Phi(\vr) &= i \left[r_{\eh}^{\lat}
\right]^{-\frac{1}{2}} \left[ s_{\rm H}(0) \Phi_{{\rm e},\vkprp,+}(\vr) + 
\Phi_{{\rm{e}},\vkprp,-}(\vr) \right]  \notag \\  & 
+i \left[r_{\eh}^{\lat} 
\right]^{\frac{1}{2}} \left[ \Phi_{{\rm h},\vkprp,-}(\vr) + s_{\rm H}(0)^* \Phi_{{\rm{h}},\vkprp,+}(\vr) \right],
\end{align}
where the scattering states $\Phi_{{\rm e},\vkprp,\pm}$ and
$\Phi_{{\rm h},\vkprp,\pm}$ are obtained from those given in Eqs.\
(\ref{eq:Phie}) and (\ref{eq:Phih}) above, but with the replacement
$e^{i k_y y} \to \sin(k_y y)$ because of the hard-wall boundary
conditions. The distance to the next bound states is of the order of
the minigap $\varepsilon_0$ or the level spacing in the normal segment
extending from the superconductor, whichever is smaller.

The bound state (\ref{eq:majorana-state-inf}) is identified as a
Majorana bound state because it is invariant under electron-hole
conjugation, i.e., $\tau_1 \Phi^\ast = \Phi$ where $\tau_1$ is the 1st
Pauli matrix in electron-hole space. Alternatively, with
$\hat{\psi}_{\uparrow,\downarrow}^\dagger,
\hat{\psi}_{\uparrow,\downarrow}$ being electron and hole creation
operators, $\Phi$ corresponds to the field operator \begin{align}
  \label{eq:gamma-maj}
  \gamma &= \int dx [u_\uparrow(\mathbf x) \hat{\psi}_\uparrow (\mathbf x)+
  v_\uparrow(\mathbf x) \hat{\psi}^\dagger_{\uparrow} (\mathbf x)]   \, ,
\end{align}
with $ u_\uparrow = \Phi_1 $ and $ v_{\uparrow} = \Phi_3 $ given by
the electron- and hole spin-up component of $\Phi$, respectively. This
operator satisfies the condition
\begin{align}
\gamma &= \gamma^\dagger,
\end{align}
which is the defining characteristic of a Majorana state. Being a
Majorana bound state, $\Phi$ is stable against perturbations because,
by particle-hole symmetry, a perturbation that moves
$\Phi$ to some finite energy $\varepsilon \neq 0$ must
generate a pair of states at $\pm \varepsilon$. Since $\Phi$ is a
single state this is not possible.

We note that there is one Majorana mode per transverse mode in the
half-metallic wire. Disorder, which is not included here, will lead to
interactions between these modes, which will cause Majorana modes to
pairwise combine into standard fermionic excitations. If the number of
transverse modes is odd, a single Majorana mode is guaranteed to
remain present at the end of the half-metallic wire.\cite{Potter2010}

If the half-metallic quantum wire has a finite length $L$, the
Majorana bound states at the two ends will interact, so that the
excitation acquires a finite energy, exponentially small in the length
of the wire. This finite excitation energy can be calculated from the
full scattering matrix ${\cal S}(\varepsilon)$ of the lateral H-S
contact, calculated in Sec.~\ref{sec:lateral-geometry}, and the
reflection amplitudes $s_{\rm H}(\varepsilon)$ and $s_{\rm
  H}'(\varepsilon)$ at the left and rights ends of the half-metallic
wire. The energy spectrum is found from the condition
\begin{align}
  \label{eq:energy-spectrum}
  \mathrm{det} \left(1 - {\cal S}_{\rm H}(\varepsilon) {\cal S}(\varepsilon) \right) = 0,
\end{align}
where
\begin{eqnarray*}
  {\cal S}_{\rm H}(\varepsilon) &=& \left( \begin{array}{cccc} s_{\rm
        H}(\varepsilon) & 0 & 0 & 0 \\ 0 & s_{\rm H}(-\varepsilon)^* &
      0 & 0 \\ 0 & 0 & s_{\rm H}'(\varepsilon) & 0 \\ 0 & 0 & 0 &
      s_{\rm H}'(-\varepsilon)^* \end{array} \right),\\ 
  {\cal S}(\varepsilon) &=& \left( \begin{array}{cccc} 0 & r_{\rm eh}^{\lat}(\varepsilon) &
      t_{\rm ee}^{\lat}(\varepsilon) & 0 \\ r_{\rm he}^{\lat}(\varepsilon) & 0 & 0 & t_{\rm hh}^{\lat}(\varepsilon) \\
      t_{\rm ee}^{\lat}(\varepsilon) & 0 & 0 & r_{\rm eh}^{\lat}(\varepsilon) \\
      0 & t_{\rm hh}^{\lat}(\varepsilon) & r_{\rm he}^{\lat}(\varepsilon) & 0 \end{array} \right).
\end{eqnarray*}
In the limit of a long contact, $|\tilde \rho_{\eh}| L \gg 1$, we then find
\begin{align}
  \label{eq:energy-spectrum-1}
  \varepsilon_{\pm} = \pm 2 |\rho_{\eh}(\varepsilon)| \hbar v_x(\varepsilon)
  e^{-|\rho_{\eh}(\varepsilon)| L} |\sin (k_{x}(\varepsilon) L )|
  \Big|_{\varepsilon = 0}  \, ,
\end{align}
where we have set $s_{\rm H} = s'_{\rm H} = -1$.
Thus, the energy splitting decreases exponentially with the contact
length (besides accidental degeneracies for integer $k_{x}(\varepsilon
=0) L / 2 \pi$). 

It is instructive to compare our calculation with the model of a
spinless one-dimensional $p$-wave superconductor,\cite{Kitaev2000}
which has been used as a phenomenological model description of the
induced superconductivity in a semiconductor wire with a strong
magnetic field and spin-orbit coupling.\cite{Lee2009-pa,Potter2010}
This model has the Hamiltonian
\begin{equation}
  H = \frac{p^2}{2 m} \tau_0 + \Delta' p \tau_1,
\end{equation}
where $\tau_0$ is the $2 \times 2$ unit matrix in electron-hole space
and $\Delta'$ the effective $p$-wave superconducting order
parameter. Comparing with our calculation, and specializing to a
quantum wire with one quantized mode only, for which $k_z = \pi/d$, we
identify
\begin{equation}
  |\Delta'| = \frac{\pi \hbar \tau(\vkprp)}{4 d^2 (k_{\rm F,S}^2 -
    k_x^2 - k_y^2)} \sqrt{\Omega^2_{1,x} + \Omega^2_{2,x}}, 
\end{equation}
where $\tau(\vkprp)$ is the transparency of the interface at the
relevant (lowest) transverse mode.

\section{Discussion and Conclusion}

In this article, we have shown that spin-orbit interaction in a
singlet superconductor gives rise to a triplet proximity effect if the
superconductor (S) is coupled to a half-metallic ferromagnet (H). We
have calculated the conductance of a HS junction and the Josephson
current of a SHS junction in both a serial geometry and in a lateral
contact geometry. Because of the coherent effect of multiple Andreev
reflections, the effective Andreev amplitudes for a lateral contact
geometry are significantly enhanced in comparison to those at a serial
geometry. In particular, multiple Andreev reflections at the interface
between a clean (disorder-free) half-metallic film or wire and a
superconductor can lead to a fully developed triplet proximity effect
in the half metal, with an Andreev reflection amplitude of unit
magnitude.

The results found here have been derived under the assumption of a
ballistic system, {\em i.e.}, without taking into account disorder
scattering in the half-metal or in the superconductor. For the
single-reflection amplitude $r_{\eh}$ in Sec.~\ref{sec:s-matrix-hm}
this does not strongly restrict the validity of the result since the
Andreev reflection amplitude is a microscopic property of the
interface: $r_{\rm he}$ is determined by matching the eigenstates on
length scales of the Fermi wave length in \HM and S and the wave
function decay length in S. If the disorder is weak, such that the
mean free path $l$ exceeds these microscopic length scales, $r_{\eh}$
will be unchanged by the presence of disorder. In this way, the
microscopic Andreev reflection amplitudes calculated in Sec.\
\ref{sec:s-matrix-hm} may also serve as a starting point for studies
of the conductance of a disordered HS junction or the Josephson
current in a disordered S-\HM-S junctions. (For example, the Josephson
current through a disordered or a chaotic Josephson junction can be
found by combining the reflection amplitudes of the clean
superconductor interface with the normal-state scattering matrix of
the junction, see, {\em e.g.} Refs.\
\onlinecite{Beenakker1991,Brouwer1997}.)

On the other hand, quantities that rely on free (phase-coherent)
propagation in \HM may change qualitatively in the presence of
disorder. Specifically, the effective Andreev reflection amplitude
$r_{\eh}^{\mathrm{\lat}}$ of the lateral contact has been obtained by
phase-coherently summing single reflection amplitudes. At the Fermi
energy, these multiple Andreev reflections add constructively because
the momentum $\vkp$ parallel to the interface is conserved and
amplitudes of subsequent reflections have the same sign. However,
scattering from impurities under the contact will lead to a summation
over single amplitudes with different incident angles. Since the
Andreev reflection amplitudes $r_{\eh}$ and $r_{\he}$ are odd in
$\vkp$, this sum may no longer be constructive. Thus, the result for
$r_{\eh}^{\mathrm{\lat}}$ is valid for an ideal, disorder-free
lateral contact only. Since the reflection properties of a lateral
junction saturate if the junction length $L \gtrsim d/|r_{\eh}|$,
where $d$ is the thickness of the half-metallic film or wire, disorder
is not expected to significantly alter our results as long as the
elastic mean free path $l \gg d/|r_{\eh}|$.

As a particularly timely application of our calculation, we connect
the Andreev reflection amplitudes $r_{\eh}$ and $r_{\he}$ calculated
here to the existence of Majorana bound states at the ends of a
ballistic half-metallic quantum wire in (lateral) contact to a
superconductor with spin-orbit coupling. This proposal for the
construction of Majorana bound states is a variation of a recent
proposal that such Majorana bound states exist at the ends of a
semiconducting wire in contact to a superconductor, where the
semiconductor has strong spin-orbit coupling and the system is placed
in a large Zeeman field, such that a the wire becomes effectively half
metallic.\cite{Oreg2010,Lutchyn2010} In our construction, the
Zeeman field is replaced by the exchange field in the half metal, and
the spin-orbit coupling is not located in the wire, but in the
superconductor. It thus avoids the necessity of a (fine-tuned) applied
magnetic field, which could negatively interfere with the
superconducting order.

We thank B. B\'eri, J.\ Kupferschmidt, D.\ Manske, O.\ Starykh, A.\
Schnyder, F.\ von Oppen, and F.\ Wilken for discussions. This work was
supported by the Alexander von Humboldt foundation.

\appendix

\section{Perturbation theory in $\hat H_{\rm SO}$}\label{sec:eigenstates-hm}

An alternative calculation of the Andreev reflection amplitudes
$r_{\eh}$ and $r_{\he}$ makes use of perturbation theory in the
spin-orbit interaction $\hat H_{\rm SO}$. For this calculation, finite
lateral dimensions $W_x \times W_y$ are assumed, with periodic
boundary conditions in the $x$ and $y$ directions.

As before, we consider wavefunctions proportional to $e^{i k_x x + i
  k_y y}$. In the absence of spin-orbit coupling, there are two
linearly independent solutions of the Bogoliubov-de Gennes equation
for each pair $k_x, k_y$. The first of these is ``electron-like'', and
of the general form
\begin{eqnarray}
  \Psi_{{\rm e}\vkp}(\vr) &=&
  \frac{c_{{\rm e}\uparrow} e^{i
      k_z(\varepsilon) z} + c_{{\rm e}\uparrow}' e^{- i k_z(\varepsilon) z}}{\sqrt{v_z(\varepsilon) W_x W_y}}
  \nonumber \\ && \mbox{} \times
   e^{i k_x x + i k_y y}  \left(1,0,0,0 \right)^{\rm T}
  \nonumber \\ && \mbox{}
  +
  c_{{\rm h}\downarrow} e^{i k_x x + i k_y y + \kappa_z(-\varepsilon) z}
  \left( 0,0,0,1 \right)^{\rm T} ~~~
\end{eqnarray}
for $z < 0$ and
\begin{eqnarray}
  \Psi_{{\rm e}\vkp}(\vr) &=&
  d_{\uparrow}' e^{i k_x x + i k_y y + i q_+ z} 
  \left( 
  1,0,0,e^{-i \phi - i \eta}  \right)^{\rm T}
  \nonumber \\ && \mbox{} +
   d_{\uparrow} e^{i k_x x + i k_y y + i q_- z}
  \left( 
  1,0,0,e^{-i \phi + i \eta}  \right)^{\rm T} ~~~
\end{eqnarray} 
for $z > 0$, where 
\begin{equation}
  q_{s} = s \sqrt{k_{{\rm F,S}}^2 - k_x^2 - k_y^2 + 2 i s m \sqrt{\Delta_0^2-\varepsilon^2}}.
\end{equation}
The boundary conditions (\ref{eq:bc-veloc}) with $\Omega_{j,z} = 0$
give four equations for the five coefficients $c_{{\rm e},\uparrow}$,
$c_{{\rm e},\uparrow}'$, $c_{{\rm h},\downarrow}$, $d_{\uparrow}$, and
$d_{\uparrow}'$, so that one coefficient can be chosen
freely. Choosing $c_{{\rm e},\uparrow} = 1$ one obtains the ``retarded
scattering state'' $\Psi_{{\rm e},\vkp}^{\rm R}$, while choosing
$c_{{\rm e},\uparrow}' = 1$ one obtains the ``advanced scattering
state'' $\Psi_{{\rm e},\vkp}^{\rm A}$.

The second scattering state is ``hole-like'' and has the general form
\begin{eqnarray}
  \Psi_{{\rm h}\vkp}(\vr) &=&
  \frac{c_{{\rm h}\uparrow}   e^{ - i
      k_z(-\varepsilon) z} + c_{{\rm h}\uparrow}'   e^{i k_z(-\varepsilon) z}}{\sqrt{v_z(-\varepsilon) W_x W_y}}
  \nonumber \\ && \mbox{}\times   e^{i k_x x + i k_y y } \, \left(0,0,1,0 \right)^{\rm T}
  \nonumber  \\ && \mbox{} +
  c_{{\rm e}\downarrow} e^{i k_x x + i k_y y + \kappa_z(\varepsilon) z}
  \left(0,1,0,0 \right)^{\rm T}
\end{eqnarray}
for $z < 0$ and
\begin{eqnarray}
  \Psi_{{\rm h}\vkp}(\vr) &=&
  d_{\downarrow}' e^{i k_x x + i k_y y + i q_+ z} 
  \left(0,1,- e^{-i \phi - i \eta},0  \right)^{\rm T}
 \\   \nonumber && \mbox{} +
  d_{\downarrow} e^{i k_x x + i k_y y + i q_- z}
  \left(  0,1,- e^{-i \phi + i \eta},0 \right)^{\rm T} 
\end{eqnarray} 
for $z > 0$. The boundary conditions (\ref{eq:bc-veloc}) with
$\Omega_{j,z} = 0$ give four equations for the five coefficients
$c_{{\rm e},\downarrow}$, $c_{{\rm h},\uparrow}$, $c_{{\rm
    h},\uparrow}'$, $d_{\downarrow}$, and $d_{\downarrow'}$. (See
Ref.\ \onlinecite{Kupferschmidt2010} for details.) Choosing $c_{{\rm
    e},\uparrow} = 1$ one obtains the ``retarded scattering state''
$\Psi_{{\rm e},\vkp}^{\rm R}$, while choosing $c_{{\rm e},\uparrow}' =
1$ one obtains the ``advanced scattering state'' $\Psi_{{\rm
    e},\vkp}^{\rm A}$.

In the Born approximation, the Andreev reflection amplitudes to first
order in the spin-orbit interaction are then found as the matrix
element
\begin{eqnarray}
  r_{\eh}(\vkp,\varepsilon) &=&
  -i \langle \Psi_{{\rm h},\vkp}^{\rm A} |
  {\cal H}_{\rm SO} |
  \Psi_{{\rm e},\vkp}^{\rm R} \rangle, 
  \label{eq:rehBorn} \\
  r_{\he}(\vkp,\varepsilon) &=&
  -i \langle \Psi_{{\rm e},\vkp}^{\rm A} |
  {\cal H}_{\rm SO} |
  \Psi_{{\rm h},\vkp}^{\rm R} \rangle,
  \label{eq:rheBorn}
\end{eqnarray}
between retarded and advanced scattering states, where ${\cal H}_{\rm
  SO}$ is given by Eq.~(\ref{eq:h-so-bdg}).

Inserting the explicit expressions for the scattering states into
Eqs.\ (\ref{eq:rehBorn}) and (\ref{eq:rehBorn}) then gives the Andreev
reflection amplitudes of Eqs.\ (\ref{eq:reh}) and (\ref{eq:rhe}).

Note that only the wavefunctions in the superconducting region $z > 0$
enter into the calculation of the Andreev reflection amplitudes. The
observation that the SOI proportional to $\Omega_{1,z}$ or
$\Omega_{2,z}$ does not give rise to Andreev reflection to first order
in the SOI then follows from the observation that the matrix elements
(\ref{eq:rehBorn}) and (\ref{eq:rheBorn}) vanish for arbitrary
coefficients $d_{\uparrow}$, $d_{\uparrow}'$, $d_{\downarrow}$, and
$d_{\downarrow}'$ if $\hat H_{\rm SO}$ contains $\Omega_{1,z}$ or
$\Omega_{2,z}$ only.

\section{Triplet pairings in S}\label{sec:triplet-pairings-s}

In this appendix we give the results for the Andreev reflection
amplitudes in the presence of finite triplet pairings in S which are
linear in momentum. The triplet proximity effect in a half-metal in
contact to a spin-triplet superconductor was also considered by Linder
et {\it al.}\cite{Linder2007} We consider a pairing of the form
\begin{align}
  \label{eq:pairings}
 \Delta_i(\vp) = 
\sum_{j=1}^3 d_{ij}  p_j  \, ,
\end{align}
where $d_{ij}$ are the components of the triplet order parameter which
is taken as a small correction to the spin singlet s-wave pairing
$\Delta_0 e^{i \phi}$ (see Sec.~\ref{sec:interf-betw-half}). In a
similar calculation as in the main text we find the Andreev reflection
amplitude $r^{\rm t}_{\eh}$ induced by $ \Delta_i(\vp)$ to first order
in $ \Delta_i(\vp)$. The amplitude $r^{\rm t}_{\eh} = r^{\rm
  t,e}_{\eh} + r^{\rm t,o}_{\eh}$ contains contributions $r^{\rm
  t,e}_{\eh}$, $r^{\rm t,o}_{\eh}$ that are even and odd in momentum,
respectively. We find
\begin{align}
  \label{eq:AR-triplet-order-p-even}
  r^{\rm t,e}_{\eh} = &   \frac{\varepsilon \tau(\theta) }{2 (\Delta_0^2
    -\varepsilon^2)} \left( d^\ast_{13} + i d^\ast_{23}  \right)   \sqrt{k_{\rm F,S}^2 - k_{F,H}^2 \sin^2
    \theta } 
\end{align}
and
\begin{align}
  \label{eq:AR-triplet-order-p-odd}
  r^{\rm t,o}_{\eh} = & \sum_{j=1}^3 k_{||,j} \frac{i \tau(\theta)}{2}
  \Bigg[ \frac{d^\ast_{1j} + id^\ast_{2j}}{(\Delta_0^2
    -\varepsilon^2)^{\frac{1}{2}}} \notag \\ & \quad  - \frac{\Delta_0^2}{2}\frac{ \left(
      d^\ast_{1j} + id^\ast_{2j} \right) + e^{-2 i  \phi} \left(
      d_{1j} + id_{2j} \right)}{(\Delta_0^2
    -\varepsilon^2)^{\frac{3}{2}}}    \Bigg]
\end{align}
for the electron-to-hole conversion amplitudes and
\begin{align}
  \label{eq:AR-triplet-order-he}
  r^{\rm t}_{\he} (\vkp , \varepsilon ) =  \left[ r^{\rm t}_{\eh}
    (-\vkp , -\varepsilon ) \right]^\ast
\end{align}
for the opposite process. Triplet pairings can be included in the
calculation of the conductance and the Josephson current by the
substitution
\begin{equation}
  r_{\eh} \to r_{\eh} + r^{\rm t}_{\eh} \, .
\end{equation}
The enhancement due to multiple Andreev reflections found in the
second part of the article for the lateral geometry does not rely on
the details of $r_{\eh}$ and is not changed by the presence of the
$\Delta_i$.  For a detailed discussion of amplitudes that are odd or
even in frequency we refer to Ref.~\onlinecite{Kupferschmidt2010}.


\begin{thebibliography}{58}
\expandafter\ifx\csname natexlab\endcsname\relax\def\natexlab#1{#1}\fi
\expandafter\ifx\csname bibnamefont\endcsname\relax
  \def\bibnamefont#1{#1}\fi
\expandafter\ifx\csname bibfnamefont\endcsname\relax
  \def\bibfnamefont#1{#1}\fi
\expandafter\ifx\csname citenamefont\endcsname\relax
  \def\citenamefont#1{#1}\fi
\expandafter\ifx\csname url\endcsname\relax
  \def\url#1{\texttt{#1}}\fi
\expandafter\ifx\csname urlprefix\endcsname\relax\def\urlprefix{URL }\fi
\providecommand{\bibinfo}[2]{#2}
\providecommand{\eprint}[2][]{\url{#2}}

\bibitem[{\citenamefont{Bergeret et~al.}(2005)\citenamefont{Bergeret, Volkov,
  and Efetov}}]{Bergeret2005}
\bibinfo{author}{\bibfnamefont{F.~S.} \bibnamefont{Bergeret}},
  \bibinfo{author}{\bibfnamefont{A.~F.} \bibnamefont{Volkov}},
  \bibnamefont{and} \bibinfo{author}{\bibfnamefont{K.~B.}
  \bibnamefont{Efetov}}, \bibinfo{journal}{Rev. Mod. Phys.}
  \textbf{\bibinfo{volume}{77}}, \bibinfo{pages}{1321} (\bibinfo{year}{2005}).

\bibitem[{\citenamefont{Buzdin}(2005)}]{Buzdin2005}
\bibinfo{author}{\bibfnamefont{A.~I.} \bibnamefont{Buzdin}},
  \bibinfo{journal}{Rev. Mod. Phys.} \textbf{\bibinfo{volume}{77}},
  \bibinfo{pages}{935} (\bibinfo{year}{2005}).

\bibitem[{\citenamefont{Andreev}(1964)}]{Andreev1964}
\bibinfo{author}{\bibfnamefont{A.}~\bibnamefont{Andreev}},
  \bibinfo{journal}{Sov. Phys. JETP} \textbf{\bibinfo{volume}{19}},
  \bibinfo{pages}{1228} (\bibinfo{year}{1964}).

\bibitem[{\citenamefont{Bergeret et~al.}(2001)\citenamefont{Bergeret, Volkov,
  and Efetov}}]{Bergeret2001}
\bibinfo{author}{\bibfnamefont{F.~S.} \bibnamefont{Bergeret}},
  \bibinfo{author}{\bibfnamefont{A.~F.} \bibnamefont{Volkov}},
  \bibnamefont{and} \bibinfo{author}{\bibfnamefont{K.~B.}
  \bibnamefont{Efetov}}, \bibinfo{journal}{Phys. Rev. Lett.}
  \textbf{\bibinfo{volume}{86}}, \bibinfo{pages}{4096} (\bibinfo{year}{2001}).

\bibitem[{\citenamefont{Kadigrobov et~al.}(2001)\citenamefont{Kadigrobov,
  Shekhter, and Jonson}}]{Kadigrobov2001}
\bibinfo{author}{\bibfnamefont{A.}~\bibnamefont{Kadigrobov}},
  \bibinfo{author}{\bibfnamefont{R.~I.} \bibnamefont{Shekhter}},
  \bibnamefont{and} \bibinfo{author}{\bibfnamefont{M.}~\bibnamefont{Jonson}},
  \bibinfo{journal}{EPL} \textbf{\bibinfo{volume}{54}},
  \bibinfo{pages}{394} (\bibinfo{year}{2001}).

\bibitem[{\citenamefont{Petrashov et~al.}(1999)\citenamefont{Petrashov, Sosnin,
  Cox, Parsons, and Troadec}}]{Petrashov1999}
\bibinfo{author}{\bibfnamefont{V.~T.} \bibnamefont{Petrashov}},
  \bibinfo{author}{\bibfnamefont{I.~A.} \bibnamefont{Sosnin}},
  \bibinfo{author}{\bibfnamefont{I.}~\bibnamefont{Cox}},
  \bibinfo{author}{\bibfnamefont{A.}~\bibnamefont{Parsons}}, \bibnamefont{and}
  \bibinfo{author}{\bibfnamefont{C.}~\bibnamefont{Troadec}},
  \bibinfo{journal}{Phys. Rev. Lett.} \textbf{\bibinfo{volume}{83}},
  \bibinfo{pages}{3281} (\bibinfo{year}{1999}).

\bibitem[{\citenamefont{Sosnin et~al.}(2006)\citenamefont{Sosnin, Cho,
  Petrashov, and Volkov}}]{Sosnin2006}
\bibinfo{author}{\bibfnamefont{I.}~\bibnamefont{Sosnin}},
  \bibinfo{author}{\bibfnamefont{H.}~\bibnamefont{Cho}},
  \bibinfo{author}{\bibfnamefont{V.~T.} \bibnamefont{Petrashov}},
  \bibnamefont{and} \bibinfo{author}{\bibfnamefont{A.~F.}
  \bibnamefont{Volkov}}, \bibinfo{journal}{Phys. Rev. Lett.}
  \textbf{\bibinfo{volume}{96}}, \bibinfo{pages}{157002}
  (\bibinfo{year}{2006}).

\bibitem[{\citenamefont{Krivoruchko and Tarenkov}(2007)}]{Krivoruchko2007}
\bibinfo{author}{\bibfnamefont{V.~N.} \bibnamefont{Krivoruchko}}
  \bibnamefont{and} \bibinfo{author}{\bibfnamefont{V.~Y.}
  \bibnamefont{Tarenkov}}, \bibinfo{journal}{Phys. Rev. B}
  \textbf{\bibinfo{volume}{75}}, \bibinfo{pages}{214508}
  (\bibinfo{year}{2007}).

\bibitem[{\citenamefont{Yates et~al.}(2007)\citenamefont{Yates, Branford,
  Magnus, Miyoshi, Morris, Cohen, Sousa, Conde, and Silvestre}}]{Yates2007}
\bibinfo{author}{\bibfnamefont{K.~A.} \bibnamefont{Yates}},
  \bibinfo{author}{\bibfnamefont{W.~R.} \bibnamefont{Branford}},
  \bibinfo{author}{\bibfnamefont{F.}~\bibnamefont{Magnus}},
  \bibinfo{author}{\bibfnamefont{Y.}~\bibnamefont{Miyoshi}},
  \bibinfo{author}{\bibfnamefont{B.}~\bibnamefont{Morris}},
  \bibinfo{author}{\bibfnamefont{L.~F.} \bibnamefont{Cohen}},
  \bibinfo{author}{\bibfnamefont{P.~M.} \bibnamefont{Sousa}},
  \bibinfo{author}{\bibfnamefont{O.}~\bibnamefont{Conde}}, \bibnamefont{and}
  \bibinfo{author}{\bibfnamefont{A.~J.} \bibnamefont{Silvestre}},
  \bibinfo{journal}{Appl. Phys. Lett.} \textbf{\bibinfo{volume}{91}},
  \bibinfo{pages}{172504} (\bibinfo{year}{2007}).

\bibitem[{\citenamefont{Khaire et~al.}(2010)\citenamefont{Khaire, Khasawneh,
  Pratt, and Birge}}]{Khaire2010}
\bibinfo{author}{\bibfnamefont{T.~S.} \bibnamefont{Khaire}},
  \bibinfo{author}{\bibfnamefont{M.~A.} \bibnamefont{Khasawneh}},
  \bibinfo{author}{\bibfnamefont{W.~P.} \bibnamefont{Pratt}}, \bibnamefont{and}
  \bibinfo{author}{\bibfnamefont{N.~O.} \bibnamefont{Birge}},
  \bibinfo{journal}{Phys. Rev. Lett.} \textbf{\bibinfo{volume}{104}},
  \bibinfo{pages}{137002} (\bibinfo{year}{2010}).

\bibitem[{\citenamefont{Wu and Samokhin}(2009)}]{Wu2009}
\bibinfo{author}{\bibfnamefont{S.}~\bibnamefont{Wu}} \bibnamefont{and}
  \bibinfo{author}{\bibfnamefont{K.~V.} \bibnamefont{Samokhin}},
  \bibinfo{journal}{Phys. Rev. B} \textbf{\bibinfo{volume}{80}},
  \bibinfo{eid}{014516} (\bibinfo{year}{2009}).

\bibitem[{\citenamefont{Linder and Sudb\o{}}(2007)}]{Linder2007}
\bibinfo{author}{\bibfnamefont{J.}~\bibnamefont{Linder}} \bibnamefont{and}
  \bibinfo{author}{\bibfnamefont{A.}~\bibnamefont{Sudb\o{}}},
  \bibinfo{journal}{Phys. Rev. B} \textbf{\bibinfo{volume}{76}},
  \bibinfo{pages}{054511} (\bibinfo{year}{2007}); J.~Linder, M.~Cuoco,
  A.~Sudb\o, ibid. \textbf{81}, 174526 (2010).

\bibitem[{\citenamefont{Keizer et~al.}(2006)\citenamefont{Keizer, Goennenwein,
  Klapwijk, Miao, Xiao, and Gupta}}]{Keizer2006}
\bibinfo{author}{\bibfnamefont{R.~S.} \bibnamefont{Keizer}},
  \bibinfo{author}{\bibfnamefont{S.~T.~B.} \bibnamefont{Goennenwein}},
  \bibinfo{author}{\bibfnamefont{T.~M.} \bibnamefont{Klapwijk}},
  \bibinfo{author}{\bibfnamefont{G.}~\bibnamefont{Miao}},
  \bibinfo{author}{\bibfnamefont{G.}~\bibnamefont{Xiao}}, \bibnamefont{and}
  \bibinfo{author}{\bibfnamefont{A.}~\bibnamefont{Gupta}},
  \bibinfo{journal}{Nature} \textbf{\bibinfo{volume}{439}},
  \bibinfo{pages}{825} (\bibinfo{year}{2006}).

\bibitem[{\citenamefont{Anwar et~al.}(2010)\citenamefont{Anwar, Czeschka,
  Hesselberth, Porcu, and Aarts}}]{Anwar2010}
\bibinfo{author}{\bibfnamefont{M.~S.} \bibnamefont{Anwar}},
  \bibinfo{author}{\bibfnamefont{F.}~\bibnamefont{Czeschka}},
  \bibinfo{author}{\bibfnamefont{M.}~\bibnamefont{Hesselberth}},
  \bibinfo{author}{\bibfnamefont{M.}~\bibnamefont{Porcu}}, \bibnamefont{and}
  \bibinfo{author}{\bibfnamefont{J.}~\bibnamefont{Aarts}},
  \bibinfo{journal}{Phys. Rev. B} \textbf{\bibinfo{volume}{82}},
  \bibinfo{pages}{100501} (\bibinfo{year}{2010}).

\bibitem[{\citenamefont{Eschrig and L\"ofwander}(2008)}]{Eschrig2008}
\bibinfo{author}{\bibfnamefont{M.}~\bibnamefont{Eschrig}} \bibnamefont{and}
  \bibinfo{author}{\bibfnamefont{T.}~\bibnamefont{L\"ofwander}},
  \bibinfo{journal}{Nature Phys.} \textbf{\bibinfo{volume}{4}},
  \bibinfo{pages}{138} (\bibinfo{year}{2008}).

\bibitem[{\citenamefont{B\'{e}ri et~al.}(2009)\citenamefont{B\'{e}ri,
  Kupferschmidt, Beenakker, and Brouwer}}]{beri2009}
\bibinfo{author}{\bibfnamefont{B.}~\bibnamefont{B\'{e}ri}},
  \bibinfo{author}{\bibfnamefont{J.~N.} \bibnamefont{Kupferschmidt}},
  \bibinfo{author}{\bibfnamefont{C.~W.~J.} \bibnamefont{Beenakker}},
  \bibnamefont{and} \bibinfo{author}{\bibfnamefont{P.~W.}
    \bibnamefont{Brouwer}}, \bibinfo{journal}{Phys. Rev. B}
  \textbf{\bibinfo{volume}{79}}, \bibinfo{eid}{024517}
  (\bibinfo{year}{2009}).

\bibitem[{\citenamefont{Kupferschmidt and Brouwer}(2010)}]{Kupferschmidt2010}
\bibinfo{author}{\bibfnamefont{J.~N.} \bibnamefont{Kupferschmidt}}
  \bibnamefont{and} \bibinfo{author}{\bibfnamefont{P.~W.}
  \bibnamefont{Brouwer}},
  \bibinfo{note}{arXiv:1009.3163}.

\bibitem[{\citenamefont{Kupferschmidt and Brouwer}(2009)}]{Kupferschmidt2009}
\bibinfo{author}{\bibfnamefont{J.~N.} \bibnamefont{Kupferschmidt}}
  \bibnamefont{and} \bibinfo{author}{\bibfnamefont{P.~W.}
  \bibnamefont{Brouwer}}, \bibinfo{journal}{Phys. Rev. B}
  \textbf{\bibinfo{volume}{80}}, \bibinfo{pages}{214537}
  (\bibinfo{year}{2009}).

\bibitem[{\citenamefont{Wilken}(2010)}]{Wilken2010}
\bibinfo{author}{\bibfnamefont{F.}~\bibnamefont{Wilken}},
  \bibinfo{type}{Diploma thesis}, \bibinfo{school}{Freie Universit\"at Berlin}
  (\bibinfo{year}{2010}).

\bibitem[{\citenamefont{LaShell et~al.}(1996)\citenamefont{LaShell, McDougall,
  and Jensen}}]{LaShell1996}
\bibinfo{author}{\bibfnamefont{S.}~\bibnamefont{LaShell}},
  \bibinfo{author}{\bibfnamefont{B.~A.} \bibnamefont{McDougall}},
  \bibnamefont{and} \bibinfo{author}{\bibfnamefont{E.}~\bibnamefont{Jensen}},
  \bibinfo{journal}{Phys. Rev. Lett.} \textbf{\bibinfo{volume}{77}},
  \bibinfo{pages}{3419} (\bibinfo{year}{1996}).

\bibitem[{\citenamefont{Rashba}(1960)}]{rashba:1960a}
\bibinfo{author}{\bibfnamefont{E.~I.} \bibnamefont{Rashba}},
  \bibinfo{journal}{Sov. Phys. Solid State} \textbf{\bibinfo{volume}{2}},
  \bibinfo{pages}{1109} (\bibinfo{year}{1960}).

\bibitem[{\citenamefont{Nitta et~al.}(1997)\citenamefont{Nitta, Akazaki,
  Takayanagi, and Enoki}}]{Nitta1997}
\bibinfo{author}{\bibfnamefont{J.}~\bibnamefont{Nitta}},
  \bibinfo{author}{\bibfnamefont{T.}~\bibnamefont{Akazaki}},
  \bibinfo{author}{\bibfnamefont{H.}~\bibnamefont{Takayanagi}},
  \bibnamefont{and} \bibinfo{author}{\bibfnamefont{T.}~\bibnamefont{Enoki}},
  \bibinfo{journal}{Phys. Rev. Lett.} \textbf{\bibinfo{volume}{78}},
  \bibinfo{pages}{1335} (\bibinfo{year}{1997}).

\bibitem[{\citenamefont{Kato et~al.}(2004)\citenamefont{Kato, Myers, Gossard,
  and Awschalom}}]{Kato2004}
\bibinfo{author}{\bibfnamefont{Y.~K.} \bibnamefont{Kato}},
  \bibinfo{author}{\bibfnamefont{R.~C.} \bibnamefont{Myers}},
  \bibinfo{author}{\bibfnamefont{A.~C.} \bibnamefont{Gossard}},
  \bibnamefont{and} \bibinfo{author}{\bibfnamefont{D.~D.}
  \bibnamefont{Awschalom}}, \bibinfo{journal}{Nature}
  \textbf{\bibinfo{volume}{427}}, \bibinfo{pages}{50} (\bibinfo{year}{2004}).

\bibitem[{\citenamefont{Bauer et~al.}(2004)\citenamefont{Bauer, Hilscher,
  Michor, Paul, Scheidt, Gribanov, Seropegin, No\"el, Sigrist, and
  Rogl}}]{Bauer2004}
\bibinfo{author}{\bibfnamefont{E.}~\bibnamefont{Bauer}},
  \bibinfo{author}{\bibfnamefont{G.}~\bibnamefont{Hilscher}},
  \bibinfo{author}{\bibfnamefont{H.}~\bibnamefont{Michor}},
  \bibinfo{author}{\bibfnamefont{C.}~\bibnamefont{Paul}},
  \bibinfo{author}{\bibfnamefont{E.~W.} \bibnamefont{Scheidt}},
  \bibinfo{author}{\bibfnamefont{A.}~\bibnamefont{Gribanov}},
  \bibinfo{author}{\bibfnamefont{Y.}~\bibnamefont{Seropegin}},
  \bibinfo{author}{\bibfnamefont{H.}~\bibnamefont{No\"el}},
  \bibinfo{author}{\bibfnamefont{M.}~\bibnamefont{Sigrist}}, \bibnamefont{and}
  \bibinfo{author}{\bibfnamefont{P.}~\bibnamefont{Rogl}},
  \bibinfo{journal}{Phys. Rev. Lett.} \textbf{\bibinfo{volume}{92}},
  \bibinfo{pages}{027003} (\bibinfo{year}{2004}).

\bibitem[{\citenamefont{Sigrist et~al.}(2007)\citenamefont{Sigrist, Agterberg,
  Frigeri, Hayashi, Kaur, Koga, Milat, Wakabayashi, and Yanase}}]{Sigrist2007}
\bibinfo{author}{\bibfnamefont{M.}~\bibnamefont{Sigrist}} et {\it al.},
  \bibinfo{journal}{J. Mag. Mag. Mat.}
  \textbf{\bibinfo{volume}{310}}, \bibinfo{pages}{536 } (\bibinfo{year}{2007}).

\bibitem[{\citenamefont{Agterberg}(2003)}]{Agterberg2003}
\bibinfo{author}{\bibfnamefont{D.~F.} \bibnamefont{Agterberg}},
  \bibinfo{journal}{Physica C}
  \textbf{\bibinfo{volume}{387}}, \bibinfo{pages}{13 } (\bibinfo{year}{2003}).

\bibitem[{\citenamefont{Samokhin}(2004)}]{Samokhin2004}
\bibinfo{author}{\bibfnamefont{K.~V.} \bibnamefont{Samokhin}},
  \bibinfo{journal}{Phys. Rev. B} \textbf{\bibinfo{volume}{70}},
  \bibinfo{pages}{104521} (\bibinfo{year}{2004}).

\bibitem[{\citenamefont{Kaur et~al.}(2005)\citenamefont{Kaur, Agterberg, and
  Sigrist}}]{Kaur2005}
\bibinfo{author}{\bibfnamefont{R.~P.} \bibnamefont{Kaur}},
  \bibinfo{author}{\bibfnamefont{D.~F.} \bibnamefont{Agterberg}},
  \bibnamefont{and} \bibinfo{author}{\bibfnamefont{M.}~\bibnamefont{Sigrist}},
  \bibinfo{journal}{Phys. Rev. Lett.} \textbf{\bibinfo{volume}{94}},
  \bibinfo{pages}{137002} (\bibinfo{year}{2005}).

\bibitem[{\citenamefont{Sigrist and Ueda}(1991)}]{Sigrist1991}
\bibinfo{author}{\bibfnamefont{M.}~\bibnamefont{Sigrist}} \bibnamefont{and}
  \bibinfo{author}{\bibfnamefont{K.}~\bibnamefont{Ueda}},
  \bibinfo{journal}{Rev. Mod. Phys.} \textbf{\bibinfo{volume}{63}},
  \bibinfo{pages}{239} (\bibinfo{year}{1991}).

\bibitem[{\citenamefont{Frigeri et~al.}(2004)\citenamefont{Frigeri, Agterberg,
  Koga, and Sigrist}}]{Frigeri2004}
\bibinfo{author}{\bibfnamefont{P.~A.} \bibnamefont{Frigeri}},
  \bibinfo{author}{\bibfnamefont{D.~F.} \bibnamefont{Agterberg}},
  \bibinfo{author}{\bibfnamefont{A.}~\bibnamefont{Koga}}, \bibnamefont{and}
  \bibinfo{author}{\bibfnamefont{M.}~\bibnamefont{Sigrist}},
  \bibinfo{journal}{Phys. Rev. Lett.} \textbf{\bibinfo{volume}{92}},
  \bibinfo{pages}{097001} (\bibinfo{year}{2004}).

\bibitem[{\citenamefont{Nelson et~al.}(2004)\citenamefont{Nelson, Mao, Maeno,
  and Liu}}]{Nelson2004}
\bibinfo{author}{\bibfnamefont{K.~D.} \bibnamefont{Nelson}},
  \bibinfo{author}{\bibfnamefont{Z.~Q.} \bibnamefont{Mao}},
  \bibinfo{author}{\bibfnamefont{Y.}~\bibnamefont{Maeno}}, \bibnamefont{and}
  \bibinfo{author}{\bibfnamefont{Y.}~\bibnamefont{Liu}},
  \bibinfo{journal}{Science} \textbf{\bibinfo{volume}{306}},
  \bibinfo{pages}{1151} (\bibinfo{year}{2004}).

\bibitem[{\citenamefont{Edelstein}(1995)}]{Edelstein1995}
\bibinfo{author}{\bibfnamefont{V.~M.} \bibnamefont{Edelstein}},
  \bibinfo{journal}{J. Phys. C} \textbf{\bibinfo{volume}{7}},
  \bibinfo{pages}{1} (\bibinfo{year}{1995}).

\bibitem[{\citenamefont{Fujimoto}(2005)}]{Fujimoto2005}
\bibinfo{author}{\bibfnamefont{S.}~\bibnamefont{Fujimoto}},
  \bibinfo{journal}{Phys. Rev. B} \textbf{\bibinfo{volume}{72}},
  \bibinfo{pages}{024515} (\bibinfo{year}{2005}).

\bibitem[{\citenamefont{Gor'kov and Rashba}(2001)}]{Gorkov2001}
\bibinfo{author}{\bibfnamefont{L.~P.} \bibnamefont{Gor'kov}} \bibnamefont{and}
  \bibinfo{author}{\bibfnamefont{E.~I.} \bibnamefont{Rashba}},
  \bibinfo{journal}{Phys. Rev. Lett.} \textbf{\bibinfo{volume}{87}},
  \bibinfo{pages}{037004} (\bibinfo{year}{2001}).

\bibitem[{\citenamefont{Yip}(2002)}]{Yip2002}
\bibinfo{author}{\bibfnamefont{S.~K.} \bibnamefont{Yip}},
  \bibinfo{journal}{Phys. Rev. B} \textbf{\bibinfo{volume}{65}},
  \bibinfo{pages}{144508} (\bibinfo{year}{2002}).

\bibitem[{\citenamefont{Braude and Nazarov}(2007)}]{Braude2007}
\bibinfo{author}{\bibfnamefont{V.}~\bibnamefont{Braude}} \bibnamefont{and}
  \bibinfo{author}{\bibfnamefont{Y.~V.} \bibnamefont{Nazarov}},
  \bibinfo{journal}{Phys. Rev. Lett.} \textbf{\bibinfo{volume}{98}},
  \bibinfo{pages}{077003} (\bibinfo{year}{2007}).

\bibitem[{\citenamefont{Kitaev}()}]{Kitaev2000}
\bibinfo{author}{\bibfnamefont{A.~Y.} \bibnamefont{Kitaev}},
  \bibinfo{note}{arXiv:cond-mat/0010440}.

\bibitem[{\citenamefont{Sau et~al.}(2010)\citenamefont{Sau, Lutchyn, Tewari,
  and Das~Sarma}}]{kn:sau2010}
\bibinfo{author}{\bibfnamefont{J.~D.} \bibnamefont{Sau}},
  \bibinfo{author}{\bibfnamefont{R.~M.} \bibnamefont{Lutchyn}},
  \bibinfo{author}{\bibfnamefont{S.}~\bibnamefont{Tewari}}, \bibnamefont{and}
  \bibinfo{author}{\bibfnamefont{S.}~\bibnamefont{Das~Sarma}},
  \bibinfo{journal}{Phys. Rev. Lett.} \textbf{\bibinfo{volume}{104}},
  \bibinfo{pages}{040502} (\bibinfo{year}{2010}).

\bibitem[{\citenamefont{Lee}()}]{Lee2009-pa}
\bibinfo{author}{\bibfnamefont{P.~A.} \bibnamefont{Lee}},
  \bibinfo{note}{arXiv0907.2681}.

\bibitem[{\citenamefont{Lutchyn et~al.}(2010)\citenamefont{Lutchyn, Sau, and
  Das~Sarma}}]{Lutchyn2010}
\bibinfo{author}{\bibfnamefont{R.~M.} \bibnamefont{Lutchyn}},
  \bibinfo{author}{\bibfnamefont{J.~D.} \bibnamefont{Sau}}, \bibnamefont{and}
  \bibinfo{author}{\bibfnamefont{S.}~\bibnamefont{Das~Sarma}},
  \bibinfo{journal}{Phys. Rev. Lett.} \textbf{\bibinfo{volume}{105}},
  \bibinfo{pages}{077001} (\bibinfo{year}{2010}).

\bibitem[{\citenamefont{Oreg et~al.}(2010)\citenamefont{Oreg, Refael, and von
  Oppen}}]{Oreg2010}
\bibinfo{author}{\bibfnamefont{Y.}~\bibnamefont{Oreg}},
  \bibinfo{author}{\bibfnamefont{G.}~\bibnamefont{Refael}}, \bibnamefont{and}
  \bibinfo{author}{\bibfnamefont{F.}~\bibnamefont{von Oppen}},
  \bibinfo{journal}{Phys. Rev. Lett.} \textbf{\bibinfo{volume}{105}},
  \bibinfo{pages}{177002} (\bibinfo{year}{2010}).

\bibitem[{\citenamefont{Potter and Lee}()}]{Potter2010}
\bibinfo{author}{\bibfnamefont{A.~C.} \bibnamefont{Potter}} \bibnamefont{and}
  \bibinfo{author}{\bibfnamefont{P.~A.} \bibnamefont{Lee}},
  \bibinfo{note}{arXiv:1007.4569}.

\bibitem[{\citenamefont{Alicea et~al.}(2010)\citenamefont{Alicea, Oreg, Refael,
  von Oppen, and Fisher}}]{Alicea2010}
\bibinfo{author}{\bibfnamefont{J.}~\bibnamefont{Alicea}},
  \bibinfo{author}{\bibfnamefont{Y.}~\bibnamefont{Oreg}},
  \bibinfo{author}{\bibfnamefont{G.}~\bibnamefont{Refael}},
  \bibinfo{author}{\bibfnamefont{F.}~\bibnamefont{von Oppen}},
  \bibnamefont{and} \bibinfo{author}{\bibfnamefont{M.~P.~A.}
  \bibnamefont{Fisher}} (\bibinfo{year}{2010}),
  \bibinfo{note}{arXiv:1006.4395}.

\bibitem[{\citenamefont{Nayak et~al.}(2008)\citenamefont{Nayak, Simon, Stern,
  Freedman, and Das~Sarma}}]{Nayak2008}
\bibinfo{author}{\bibfnamefont{C.}~\bibnamefont{Nayak}},
  \bibinfo{author}{\bibfnamefont{S.~H.} \bibnamefont{Simon}},
  \bibinfo{author}{\bibfnamefont{A.}~\bibnamefont{Stern}},
  \bibinfo{author}{\bibfnamefont{M.}~\bibnamefont{Freedman}}, \bibnamefont{and}
  \bibinfo{author}{\bibfnamefont{S.}~\bibnamefont{Das~Sarma}},
  \bibinfo{journal}{Rev. Mod. Phys.} \textbf{\bibinfo{volume}{80}},
  \bibinfo{pages}{1083} (\bibinfo{year}{2008}).

\bibitem[{\citenamefont{Likharev}(1979)}]{Likharev1979}
\bibinfo{author}{\bibfnamefont{K.~K.} \bibnamefont{Likharev}},
  \bibinfo{journal}{Rev. Mod. Phys.} \textbf{\bibinfo{volume}{51}},
  \bibinfo{pages}{101} (\bibinfo{year}{1979}).

\bibitem{Note1}%
  \bibinfo {note} {Note that the SOI is position dependent and the operator
  ordering of $z$ and $p_z$ in $\protect \mathaccentV {hat}05EH_{\protect \rm
  SO}$ has to be defined. Only for linear-in-momentum SOIs the ordering is
  uniquely determined by hermiticity of the Hamiltonian. SOIs with higher
  orders in $p_z$ require a microscopic derivation of an effective-mass
  Hamiltonian for an atomically sharp interface, see, {\protect \em e.g.}, M.\
  G.\ Burt, Phys.\ Rev.\ B {\protect \bf 50}, 7518 (1994).}%

\bibitem[{\citenamefont{Tinkham}(2004)}]{Tinkham2004}
\bibinfo{author}{\bibfnamefont{M.}~\bibnamefont{Tinkham}},
  \emph{\bibinfo{title}{Introduction to Superconductivity}}
  (\bibinfo{publisher}{Dover Publications}, \bibinfo{year}{2004}).

\bibitem[{\citenamefont{Blonder et~al.}(1982)\citenamefont{Blonder, Tinkham,
  and Klapwijk}}]{Blonder1982}
\bibinfo{author}{\bibfnamefont{G.~E.} \bibnamefont{Blonder}},
  \bibinfo{author}{\bibfnamefont{M.}~\bibnamefont{Tinkham}}, \bibnamefont{and}
  \bibinfo{author}{\bibfnamefont{T.~M.} \bibnamefont{Klapwijk}},
  \bibinfo{journal}{Phys. Rev. B} \textbf{\bibinfo{volume}{25}},
  \bibinfo{pages}{4515} (\bibinfo{year}{1982}).

\bibitem[{\citenamefont{Takane and Ebisawa}(1992)}]{Takane1992}
\bibinfo{author}{\bibfnamefont{Y.}~\bibnamefont{Takane}} \bibnamefont{and}
  \bibinfo{author}{\bibfnamefont{H.}~\bibnamefont{Ebisawa}},
  \bibinfo{journal}{J. Phys. Soc. Japan}
  \textbf{\bibinfo{volume}{61}}, \bibinfo{pages}{1685} (\bibinfo{year}{1992}).

\bibitem[{\citenamefont{Beenakker}(1991)}]{Beenakker1991}
\bibinfo{author}{\bibfnamefont{C.~W.~J.} \bibnamefont{Beenakker}},
  \bibinfo{journal}{Phys. Rev. Lett.} \textbf{\bibinfo{volume}{67}},
  \bibinfo{pages}{3836} (\bibinfo{year}{1991}).

\bibitem[{\citenamefont{Brouwer and Beenakker}(1997)}]{Brouwer1997}
\bibinfo{author}{\bibfnamefont{P.}~\bibnamefont{Brouwer}} \bibnamefont{and}
  \bibinfo{author}{\bibfnamefont{C.}~\bibnamefont{Beenakker}},
  \bibinfo{journal}{Chaos, Solitons \& Fractals} \textbf{\bibinfo{volume}{8}},
  \bibinfo{pages}{1249 } (\bibinfo{year}{1997}).

\bibitem[{\citenamefont{Ishii}(1970)}]{Ishii1970}
\bibinfo{author}{\bibfnamefont{C.}~\bibnamefont{Ishii}},
  \bibinfo{journal}{Prog. Theoret. Phys.}
  \textbf{\bibinfo{volume}{44}}, \bibinfo{pages}{1525} (\bibinfo{year}{1970}).

\bibitem[{\citenamefont{Fu and Kane}(2009)}]{kn:fu2009b}
\bibinfo{author}{\bibfnamefont{L.}~\bibnamefont{Fu}} \bibnamefont{and}
  \bibinfo{author}{\bibfnamefont{C.~L.} \bibnamefont{Kane}},
  \bibinfo{journal}{Phys. Rev. B} \textbf{\bibinfo{volume}{79}},
  \bibinfo{pages}{161408} (\bibinfo{year}{2009}).

\bibitem[{\citenamefont{Moore and Read}(1991)}]{kn:moore1991}
\bibinfo{author}{\bibfnamefont{G.}~\bibnamefont{Moore}} \bibnamefont{and}
  \bibinfo{author}{\bibfnamefont{N.}~\bibnamefont{Read}},
  \bibinfo{journal}{Nucl. Phys. B} \textbf{\bibinfo{volume}{360}},
  \bibinfo{pages}{362} (\bibinfo{year}{1991}).

\bibitem[{\citenamefont{Read and Green}(2000)}]{kn:read2000}
\bibinfo{author}{\bibfnamefont{N.}~\bibnamefont{Read}} \bibnamefont{and}
  \bibinfo{author}{\bibfnamefont{D.}~\bibnamefont{Green}},
  \bibinfo{journal}{Phys. Rev. B} \textbf{\bibinfo{volume}{61}},
  \bibinfo{pages}{10267} (\bibinfo{year}{2000}).

\bibitem[{\citenamefont{Ivanov}(2001)}]{kn:ivanov2001}
\bibinfo{author}{\bibfnamefont{D.~A.} \bibnamefont{Ivanov}},
  \bibinfo{journal}{Phys. Rev. Lett.} \textbf{\bibinfo{volume}{86}},
  \bibinfo{pages}{268} (\bibinfo{year}{2001}).

\bibitem[{\citenamefont{Stern et~al.}(2004)\citenamefont{Stern, von Oppen, and
  Mariani}}]{kn:stern2004}
\bibinfo{author}{\bibfnamefont{A.}~\bibnamefont{Stern}},
  \bibinfo{author}{\bibfnamefont{F.}~\bibnamefont{von Oppen}},
  \bibnamefont{and} \bibinfo{author}{\bibfnamefont{E.}~\bibnamefont{Mariani}},
  \bibinfo{journal}{Phys. Rev. B} \textbf{\bibinfo{volume}{70}},
  \bibinfo{pages}{205338} (\bibinfo{year}{2004}).

\bibitem[{\citenamefont{Stone and Chung}(2006)}]{kn:stone2006}
\bibinfo{author}{\bibfnamefont{M.}~\bibnamefont{Stone}} \bibnamefont{and}
  \bibinfo{author}{\bibfnamefont{S.-B.} \bibnamefont{Chung}},
  \bibinfo{journal}{Phys. Rev. B} \textbf{\bibinfo{volume}{73}},
  \bibinfo{pages}{014505} (\bibinfo{year}{2006}).

\bibitem[{\citenamefont{Fu and Kane}(2008)}]{Fu2008}
\bibinfo{author}{\bibfnamefont{L.}~\bibnamefont{Fu}} \bibnamefont{and}
  \bibinfo{author}{\bibfnamefont{C.~L.} \bibnamefont{Kane}},
  \bibinfo{journal}{Phys. Rev. Lett.} \textbf{\bibinfo{volume}{100}},
  \bibinfo{pages}{096407} (\bibinfo{year}{2008}).

\end{thebibliography}

\end{document}